\documentclass[12pt]{article}
\begin{document}
\setcounter{page}{1}
\def\theequation{\arabic{section}.\arabic{equation}}
\def\theequation{\thesection.\arabic{equation}}
\setcounter{section}{0}

\title{On the path integral representation for Wilson loops and the
non--Abelian Stokes theorem {\bf II}}

\author{M. Faber\thanks{E--mail: faber@kph.tuwien.ac.at, Tel.:
+43--1--58801--14261, Fax: +43--1--58801--14299} ,
A. N. Ivanov\thanks{E--mail: ivanov@kph.tuwien.ac.at, Tel.:
+43--1--58801--14261, Fax: +43--1--58801--14299}~$^{\ddagger}$ , N. I.
Troitskaya\thanks{Permanent Address: State
Technical University, Department of Nuclear Physics, 195251
St. Petersburg, Russian Federation}}

\date{\today}

\maketitle

\begin{center}
{\it Institut f\"ur Kernphysik, Technische Universit\"at Wien, \\ 
Wiedner Hauptstr. 8--10, A--1040 Vienna, Austria}
\end{center}

\begin{abstract}
This paper is a revised version of our recent publication Faber {\it
et al.}, Phys. Rev. D62 (2000) 025019, hep--th/9907048. The main
revision concerns the expansion into group characters that we have
used for the evaluation of path integrals over gauge degrees of
freedom. In the present paper we apply the expansion recommended by
Diakonov and Petrov in hep--lat/0008004. Our former expansion was
approximate and in the region of the particular values of parameters
violated the completeness condition by 1.4$\%$. We show that by using
the expansion into characters recommended by Diakonov and Petrov in
hep--lat/0008004 our previous results are retained and the path
integral over gauge degrees of freedom for Wilson loops derived by
Diakonov and Petrov (Phys. Lett. B224 (1989) 131 and, correspondingly,
hep--lat/0008004) by using a special regularization is erroneous and
predicts zero value for the Wilson loop. This property is obtained by
direct evaluation of path integrals for Wilson loops defined for pure
$SU(2)$ gauge fields and $Z(2)$ center vortices with spatial azimuthal
symmetry. Further we show that both derivations given by Diakonov and
Petrov for their regularized path integral, if done correctly, predict
also zero value for Wilson loops. Therefore, the application of their
path integral representation of Wilson loops cannot give {\it a new
way to check confinement in lattice} as has been declared by Diakonov
and Petrov (Phys. Lett. B242 (1990) 425 and hep--lat/0008004). Our
statement pointing out that none non--Abelian Stokes theorem can exist
for Wilson loops except the old--fashioned one derived by means of the
path-ordering procedure is retained. It is based on well--defined
properties of group characters and is not related to whatever explicit
method of evaluation of path integrals is applied. Comments on the
paper hep--lat/0008004 by Diakonov and Petrov are given. Some
misprints in our paper Phys. Rev. D62 (2000) 025019 are corrected.
\end{abstract}

\begin{center}
\end{center}

\newpage

\section{Introduction}
\setcounter{equation}{0}

\hspace{0.2in} The hypothesis of quark confinement, bridging the
hypothesis of the existence of quarks and the failure of the detection
of quarks as isolated objects, is a challenge for QCD. As a criterion
of colour confinement in QCD, Wilson [1] suggested to consider the
average value of an operator
\begin{eqnarray}\label{label1.1}
W(C) = \frac{1}{N}\,{\rm tr}\,{\cal P}_C\,e^{\textstyle i\,g\,\oint_C d
x_{\mu}\,A_{\mu}(x)} = \frac{1}{N}\,{\rm tr}\, U(C_{x x}),
\end{eqnarray}
defined on an closed loop $C$, where $A_{\mu}(x) = t^a\,A^a_{\mu}(x)$
is a gauge field, $t^a$ ($a = 1,\ldots, N^2-1$) are the generators of
the $SU(N)$ gauge group in fundamental representation normalized by
the condition ${\rm tr}\,(t^a t^b) = \delta^{ab}/2$, $g$ is the gauge
coupling constant and ${\cal P}_C$ is the
operator ordering colour matrices along the path $C$. 
The trace in Eq.(\ref{label1.1}) is computed over colour
indices. The operator
\begin{eqnarray}\label{label1.2}
U(C_{y x}) = {\cal P}_{C_{y x}}e^{\textstyle i\,g\,\int_{C_{y x}} d
z_{\mu}\,A_{\mu}(z)},
\end{eqnarray}
makes a parallel transport along the path $C_{y x}$ from $x$ to $y$. For
Wilson loops the contour $C$ defines a closed path $C_{x
x}$.  For determinations of the parallel transport
operator $U(C_{y x})$ the action of the path--ordering operator ${\cal
P}_{C_{y x}}$ is defined by the following limiting procedure [2]
\begin{eqnarray}\label{label1.3}
\hspace{-0.5in}&&U(C_{y x})={\cal P}_{C_{xy}}e^{\textstyle
i\,g\,\int_{C_{y x}} d z_{\mu}\,A_{\mu}(z)}= \lim_{n \to
\infty}\prod^{n}_{k = 1}U(C_{x_k x_{k-1}}) =\nonumber\\
\hspace{-0.5in}&&= \lim_{n \to \infty}U(C_{y x_{n -
1}})\,\ldots\,U(C_{x_2 x_1})\,U(C_{x_1 x})= \lim_{n \to \infty}
\prod^{n}_{k = 1}e^{\textstyle i\,g\,(x_k-x_{k-1})\cdot A(x_{k-1})},
\end{eqnarray}
where $C_{x_k x_{k-1}}$ is an infinitesimal segment of the path $C_{y
x}$ with $x_{0} = x$ and $x_{n} = y$. The parallel transport operator
$U(C_{x_k x_{k-1}})$ for an infinitesimal segment $C_{x_k x_{k-1}}$ is
defined by [2]:
\begin{eqnarray}\label{label1.4}
\hspace{-0.2in}U(C_{x_k x_{k-1}})= e^{\textstyle i\,g\,\int_{C_{x_k
x_{k-1}}} d z_{\mu}\,A_{\mu}(z)}=e^{\textstyle i\,g\,(x_k -
x_{k-1})\cdot A(x_{k-1})}.
\end{eqnarray}
In accordance with the definition of the path--ordering procedure
(\ref{label1.3}) the parallel transport operator $U(C_{y x})$ has the
property
\begin{eqnarray}\label{label1.5}
U(C_{y x}) = U(C_{y x_1})\,U(C_{x_1 x}),
\end{eqnarray}
where $x_1$ belongs to the path $C_{y x}$. Under gauge transformations
with a gauge function $\Omega(z)$,
\begin{eqnarray}\label{label1.6}
A_{\mu}(z) \to A^{\Omega}_{\mu}(z) = \Omega(z) A_{\mu}(z)
\Omega^{\dagger}(z) + \frac{1}{ig}\,\partial_{\mu}\Omega(z) 
\Omega^{\dagger}(z),
\end{eqnarray}
the operator $U(C_{y x})$ has a very simple transformation law
\begin{eqnarray}\label{label1.7}
U(C_{y x}) \to U^{\Omega}(C_{y x}) = \Omega(y)\,U(C_{y x})
\,\Omega^{\dagger}(x).
\end{eqnarray}
We would like to stress that this equation is valid even if the gauge
functions $\Omega(x)$ and $\Omega(y)$ differ significantly for
adjacent points $x$ and $y$.

As has been postulated by Wilson [1] the average value of the Wilson
loop $<W(C)>$ in the confinement regime should show area--law falloff
[1]
\begin{eqnarray}\label{label1.8}
<W(C)> \sim e^{\textstyle - \sigma\,{\cal A}},
\end{eqnarray}
where $\sigma$ and ${\cal A}$ are the string tension and the minimal
area of the loop, respectively. As usually the minimal area is a
rectangle of size $L\times T$.  In this case the exponent $\sigma
{\cal A}$ can be represented in the equivalent form $\sigma\,{\cal A}
= V(L)\,T$, where $V(L) = \sigma L$ is the interquark potential and
$L$ is the relative distance between quark and anti--quark.

The paper is organized as follows. In Sect.\,2 we discuss the path
integral representation for Wilson loops by using well--known
properties of group characters. The discretized form of this path
integral is naturally provided by properties of group characters and
does not need any artificial regularization. We derive a closed
expression for Wilson loops in irreducible representation $j$ of
$SU(2)$. In Sect.\,3 we extend the path integral representation to the
gauge group $SU(N)$. As an example, we give an explicit representation
for Wilson loops in the fundamental representation of $SU(3)$. In
Sects.\,4 and 5 we evaluate the path integral for Wilson loops,
suggested in Ref.[3], for two specific gauge field configurations (i)
a pure gauge field in the fundamental representation of $SU(2)$ and
(ii) $Z(2)$ center vortices with spatial azimuthal symmetry,
respectively. We show that this path integral representation fails to
describe the original Wilson loop for both cases. In Sect.\,6 we show
that the regularized evolution operator in Ref.[3] representing Wilson
loops in the form of the path integral over gauge degrees of freedom
has been evaluated incorrectly by Diakonov and Petrov. The correct
value for the evolution operator is zero. This result agrees with
those obtained in Sects.\,4 and 5. In Sect.\,7 we criticize the
removal of the oscillating factor from the evolution operator
suggested in Ref.[3] via a shift of energy levels of the
axial--symmetric top. We show that such a removal is prohibited. It
leads to a change of symmetry of the starting system from $SU(2)$ to
$U(2)$. Keeping the oscillating factor one gets a vanishing value of
Wilson loops in agreement with our results in Sects.\,4, 5 and 6. In
the Conclusion we discuss the obtained results. In Appendix we give
comments on the paper hep--lat/0008004 by Diakonov and Petrov.

\section{Path integral representation for Wilson loops}
\setcounter{equation}{0}

\hspace{0.2in} Attempts to derive a path integral representation for
Wilson loops (\ref{label1.1}), where the path ordering operator is
replaced by a path integral, have been undertaken in Refs.[3--5]. The
path integral representations have been derived for Wilson loops in
terms of gauge degrees of freedom (bosonic variables) [3,4] and
fermionic degrees of freedom (Grassmann variables) [5]. For the
derivation of the quoted path integral representations for Wilson loop
different mathematical machineries have been used. Below we discuss
the derivation of the path integral representation for Wilson loops in
terms of gauge degrees of freedom by using well--known properties of
group characters. In this case a discretized form of path integrals is
naturally provided by the properties of group characters and the
completeness condition of gauge functions. It coincides with the
standard discretization of Feynman path integrals [6] and does not
need any artificial regularization.

We argue that the path integral representation for Wilson loops
 suggested by Diakonov and Petrov in Ref.[3] is erroneous. For the
 derivation of this path integral representation Diakonov and Petrov
 have used a special regularization drawing an analogy with an
 axial--symmetric top. The moments of inertia of this top are taken
 finally to zero.  As we show below this path integral amounts to zero
 for Wilson loops defined for $SU(2)$. Therefore, it is not a surprise
 that the application of this erroneous path integral representation
 to the evaluation of the average value of Wilson loops has led to the
 conclusion that for large loops the area--law falloff is present for
 colour charges taken in any irreducible representation $r$ of $SU(N)$
 [7].  This statement has not been supported by numerical simulations
 within lattice QCD [8]. As has been verified, e.g. in Ref.[8] for
 $SU(3)$, in the confined phase and at large distances, colour charges
 with non--zero $N$--ality have string tensions of the corresponding
 fundamental representation, whereas colour charges with zero
 $N$--ality are screened by gluons and cannot form a string at large
 distances. Hence, the results obtained in Ref.[7] cannot give {\it a
 new way to check confinement in lattice} as has been declared by
 Diakonov and Petrov.

For the derivation of Wilson loops in the form of a path integral
 over gauge degrees of freedom by using well--known properties of 
 group characters it is convenient to represent $W(C)$ in terms of 
characters of irreducible representations of $SU(N)$ [9--11]
\begin{eqnarray}\label{label2.1}
W_r(C) = \frac{1}{\displaystyle  d_r}\,\chi[U_r(C_{x x})],
\end{eqnarray}
where the matrix $U_r(C_{x x})$ realizes an irreducible and
$d_r$--dimensional matrix representation $r$ of
the group $SU(N)$ with the character $\chi[U_r(C_{x x})] = {\rm
tr}[U_r(C_{x x})]$.  

In order to introduce the path integral over gauge degrees of freedom we
suggest to use 
\begin{eqnarray}\label{label2.2}
\int D\Omega_r\chi[U_r\Omega^{\dagger}_r]\,\chi[\Omega_rV_r]  =
\frac{1}{\displaystyle d_r}\,\chi[U_rV_r],
\end{eqnarray}
where the matrices $U_r$ and $V_r$ belong to the irreducible
representation $r$, and $D\Omega_r$ is the Haar measure normalized to
unity $\int D\Omega_r = 1$. The orthogonality relation for gauge
functions $\Omega_r$ reads
\begin{eqnarray}\label{label2.3}
\int D\Omega_r{(\Omega^{\dagger}_r)}_{\displaystyle a_1 b_1}
{(\Omega_r)}_{\displaystyle a_2 b_2} = \frac{1}{\displaystyle
d_r}\,\delta_{\displaystyle a_1 b_2}\,\delta_{\displaystyle b_1 a_2}.
\end{eqnarray}
By using the orthogonality relation it is convenient to represent the
Wilson loop in the form of the integral
\begin{eqnarray}\label{label2.4}
W_r(C) = \frac{1}{\displaystyle d_r}\,\int
D\Omega_r(x)\chi[\Omega_r(x)U_r(C_{x
x})\Omega^{\dagger}_r(x)].
\end{eqnarray}
According to Eq.(\ref{label1.3}) and Eq.(\ref{label1.5}) the matrix
$U_r(C_{x x})$ can be decomposed in
\begin{eqnarray}\label{label2.5}
U_r(C_{x x}) =\lim_{n\to \infty} U_r(C_{x x_{n-1}})
U_r(C_{x_{n-1} x_{n-2}})\ldots U_r(C_{x_2
x_1})U_r(C_{x_1 x}).
\end{eqnarray}
Substituting Eq.(\ref{label2.5}) in Eq.(\ref{label2.4}) and applying
$(n-1)$--times Eq.(\ref{label2.2}) we end up with
\begin{eqnarray}\label{label2.6}
W_r(C)&=& \frac{1}{d^2_r}\,\lim_{n\to \infty}\int\ldots\int D
\Omega_r(x_1) \ldots \Omega_r(x_n)\,d_r\chi[\Omega_r(x_n)U_r(C_{x_n
x_{n-1}}) \Omega^{\dagger}_r(x_{n-1})]\nonumber\\ &&\ldots
d_r\chi[\Omega_r(x_1) U_r(C_{x_1 x_n}) \Omega^{\dagger}_r(x_n)].
\end{eqnarray}
Using relations $\Omega_r(x_k) U_r(C_{x_k x_{k-1}})
\Omega^{\dagger}_r(x_{k-1}) = U^{\Omega}_r(C_{x_k x_{k-1}})$ we get
\begin{eqnarray}\label{label2.7}
W_r(C)=\frac{1}{d^2_r}\,\lim_{n\to \infty}\int \ldots \int D
\Omega_r(x_1)\ldots D \Omega_r(x_n)\, d_r\chi[U^{\Omega}_r(C_{x_n
x_{n-1}})]\ldots d_r\chi[U^{\Omega}_r(C_{x_1 x_n})].
\end{eqnarray}
The integrations over $\Omega_r(x_k)\,$ ($k=1,\ldots,n$) are well
defined. These are  standard integrations on the compact Lie group
$SU(N)$.

We should emphasize that the integrations over $\Omega_r(x_k)\,$
($k=1,\ldots,n$) are not correlated and should be carried out
independently. 

Due to Eq.(\ref{label2.2}) the discretization of Wilson loops given by
Eqs.(\ref{label2.6}) and (\ref{label2.7}) reproduces the standard
discretization of Feynman path integrals [6] where infinitesimal time
steps can be described by a classical motion. Therefore, the
discretized expression (\ref{label2.7}) can be represented formally by
\begin{eqnarray}\label{label2.8}
W_r(C) = \frac{1}{d^2_r}\int \prod_{x\in C} \left[ d_r\,D
 \Omega_r(x)\right] \, \chi[U^{\Omega}_r(C_{x x})].
\end{eqnarray}
Conversely the evaluation of this path integral corresponds to the
discretization given by Eqs.(\ref{label2.6}) and (\ref{label2.7}).
The measure of the integration over $\Omega_r(x)$ is well defined and
normalized to unity
\begin{eqnarray}\label{label2.9}
\int \prod_{x\in C} D \Omega_r(x) = \lim_{n\to \infty}
\int D \Omega_r(x_n)\int D \Omega_r(x_{n - 1})\ldots 
\int D \Omega_r(x_1) = 1.
\end{eqnarray}
Thus, for the determination of the path integral over gauge degrees of
freedom (\ref{label2.8}) we do not need to use any regularization,
since the discretization given by Eqs.(\ref{label2.6}) and
(\ref{label2.7}) are well defined.

We would like to emphasize that Eq.(\ref{label2.8}) is a continuum
analogy of the lattice version of the path integral over gauge degrees
of freedom for Wilson loops used in Eq.(2.13) of Ref.[11] for the evaluation of the average value of Wilson loops in connection with $Z(2)$ center vortices. 

Now let us to proceed to the evaluation of the characters
$\chi[U^{\Omega}_r(C_{x_k x_{k-1}})]$.  Due to the infinitesimality of
the segments $C_{x_k x_{k-1}}$ we can omit the path ordering operator
in the definition of $U^{\Omega}_r(C_{x_k x_{k-1}})$ [2]. This allows
us to evaluate the character $\chi[U^{\Omega}_r(C_{x_k x_{k-1}})]$
with $U^{\Omega}_r(C_{x_k x_{k-1}})$ taken in the form [2]
\begin{eqnarray}\label{label2.10}
 U^{\Omega}_r(C_{x_k x_{k-1}}) = \exp ig\int_{C_{x_k
 x_{k-1}}}dx_{\mu}A^{\Omega}_{\mu}(x).
\end{eqnarray}
Of course, the relation given by Eq.(\ref{label2.10}) is only defined
in the sense of a meanvalue over an infinitesimal segment $C_{x_k
x_{k-1}}$. Therefore, it can be regarded to some extent as a
smoothness condition.  Unlike the smoothness condition used by
Diakonov and Petrov [3] Eq.(\ref{label2.10}) does not corrupt the
Wilson loop represented by the path integral over the gauge degrees of
freedom.
 
The evaluation of the characters of $U^{\Omega}_r(C_{x_k x_{k-1}})$
given by Eq.(\ref{label2.10}) runs as follows. First let us consider
the simplest case, the $SU(2)$ gauge group, where we have 
$r = j = 0,1/2,1,\ldots$ and $d_j = 2j + 1$. The
character $\chi [U^{\Omega}_j(C_{x_k x_{k-1}})]$ is equal to [9,10,12]
\begin{eqnarray}\label{label2.11}
\chi [U^{\Omega}_j(C_{x_k x_{k-1}})] &=& \sum^{j}_{m_j =-j}
<m_j|U^{\Omega}_j(C_{x_k x_{k-1}})|m_j> = \nonumber\\
&=&\sum^{j}_{m_j =-j} e^{\textstyle 
i\,m_j\, \Phi[C_{x_k x_{k-1}}; A^{\Omega}]},
\end{eqnarray}
where $m_j$ is the magnetic colour quantum number, $|m_j>$ and
$m_j\,\Phi [C_{x_k x_{k-1}}; A^{\Omega}]$ are the eigenstates and
eigenvalues of the operator
\begin{eqnarray}\label{label2.12}
\hat{\Phi}[C_{x_k x_{k-1}}; A^{\Omega}] = g\int_{C_{x_k
x_{k-1}}}dx_{\mu}A^{\Omega}_{\mu}(x),
\end{eqnarray}
i.e. $\hat{\Phi}[C_{x_k x_{k-1}};
A^{\Omega}]\,|m_j>\,=\,m_j\,\Phi[C_{x_k x_{k-1}};
A^{\Omega}]\,|m_j(x_{k-1})>$. The standard procedure for the
evaluation of the eigenvalues gives $\Phi[C_{x_k x_{k-1}};
A^{\Omega}]$ in the form
\begin{eqnarray}\label{label2.13}
\Phi[C_{x_k x_{k-1}}; A^{\Omega}] = g \int_{C_{x_k
x_{k-1}}}\sqrt{\displaystyle g_{\mu\nu}[A^{\Omega}](x)
dx_{\mu}dx_{\nu}},
\end{eqnarray}
where the metric tensor can be given formally by the expression
\begin{eqnarray}\label{label2.14}
g_{\mu\nu}[A^{\Omega}](x) =
2\,{\rm tr} [A^{\Omega}_{\mu}A^{\Omega}_{\nu}](x).
\end{eqnarray}
In order to find an explicit expression for the metric tensor we
should fix a gauge. As an example let us take the Fock--Schwinger
gauge
\begin{eqnarray}\label{label2.15}
x_{\mu}A_{\mu}(x) = 0.
\end{eqnarray}
In this case the gauge field $A_{\mu}(x)$ can be expressed in terms of
the field strength tensor $G_{\mu\nu}(x)$ as follows
\begin{eqnarray}\label{label2.16}
A_{\mu}(x) = \int\limits^1_0 ds\, s\, x_{\alpha}\, G_{\alpha\mu}(x s).
\end{eqnarray}
This can be proven by using the obvious relation
\begin{eqnarray}\label{label2.17}
\hspace{-0.2in}x_{\alpha} G_{\alpha\mu}(x)&=& 
x_{\alpha} \partial_{\alpha}A_{\mu}(x)
- x_{\alpha}\partial_{\mu}A_{\alpha}(x) - i g
[x_{\alpha}A_{\alpha}(x),A_{\mu}(x)] =\nonumber\\
&=& A_{\mu}(x) +
x_{\alpha}\frac{\partial}{\partial x_{\alpha}}A_{\mu}(x),
\end{eqnarray}
valid for the Fock--Schwinger gauge $x_{\alpha}A_{\alpha}(x) = 0$.
 Replacing $x \to x s$ we can represent the r.h.s. of
 Eq.(\ref{label2.17}) as a total derivative with respect to $s$
\begin{eqnarray}\label{label2.18}
s x_{\alpha} G_{\alpha\mu}(x s) = A_{\mu}(x s) +
x_{\alpha}\frac{\partial}{\partial x_{\alpha}}A_{\mu}(x s) = \frac{d}{d
s}[s A_{\mu}(x s)].
\end{eqnarray}
Integrating out $s \in [0,1]$ we arrive at Eq.(\ref{label2.16}).

Using Eq.(\ref{label2.16}) we obtain the metric tensor
$g_{\mu\nu}[A^{\Omega}](x)$ in the form
\begin{eqnarray}\label{label2.19}
&& g_{\mu\nu}[A^{\Omega}](x) =  2 x_{\alpha} x_{\beta}
\int\limits^1_0
\int\limits^1_0 ds ds' s s' 
{\rm tr}[G^{\Omega}_{\alpha\mu}(x
s)G^{\Omega}_{\beta\nu}(x s'\,)]=\nonumber\\ 
&&= 2 x_{\alpha}
x_{\beta}\int\limits^1_0 \int\limits^1_0 ds ds' s s' 
{\rm tr}[\Omega(x
s)G_{\alpha\mu}(x s)\Omega^{\dagger}(x s)
\Omega(x s'\,)G_{\beta\nu}(x
s'\,)\Omega^{\dagger}(x s'\,)].
\end{eqnarray}
For the derivation of Eq.(\ref{label2.19}) we define the operator
$\Phi[C_{x_k x_{k-1}}; A^{\Omega}]$ of Eq.(\ref{label2.12}) following the
definition of the phase of the parallel transport operator $U(C_{x_k
x_{k-1}})$ given by Eq.(\ref{label1.4}) [2]
\begin{eqnarray}\label{label2.20}
\hat{\Phi}[C_{x_k x_{k-1}}; A^{\Omega}] &=& g\int_{C_{x_k
x_{k-1}}}dx_{\mu}A^{\Omega}_{\mu}(x)=(x_k - x_{k-1})_{\mu}
A^{\Omega}_{\mu}(x_{k-1})=\nonumber\\ &=&(x_k -
x_{k-1})_{\mu}\int\limits^1_0 ds s\,x^{\alpha}_{k-1}
\,G^{\Omega}_{\alpha\mu}(x_{k-1} s).
\end{eqnarray}
The parameter $s$ is to some extent an order parameter distinguishing
the gauge functions $\Omega(x_k)$ and $\Omega(x_{k-1})$ entering
the relation $\Omega(x_k)U(C_{x_k x_{k-1}})\Omega^{\dagger}(x_{k-1}) =
U^{\Omega}(C_{x_k x_{k-1}})$. 

Substituting Eq.(\ref{label2.11}) in Eq.(\ref{label2.7}) we arrive at
the expression for Wilson loops defined for $SU(2)$ 
\begin{eqnarray}\label{label2.21}
\hspace{-0.3in}&&W_j(C)
=\frac{1}{(2j+1)^2}\lim_{n\to\infty}\nonumber\\ &&\int D
\Omega_j(x_n)\,(2j+1)\sum^{j}_{m^{(n)}_j=-j}\!\!\!e^{\textstyle i\,
 g\, m^{(n)}_j
\int_{C_{x_1
x_n}}\sqrt{g_{\mu\nu}[A^{\Omega}](x)\,dx_{\mu}dx_{\nu}}}\nonumber\\
\hspace{-0.3in}&&\int D \Omega_j(x_{n-1})\,(2j+1)\sum^{j}_{m^{(n-1)}_j
= -j}\!\!\!  e^{\textstyle i\, g\, m^{(n-1)}_j \int_{C_{x_n
x_{n-1}}}\sqrt{g_{\mu\nu}[A^{\Omega}](x)\,dx_{\mu}dx_{\nu}}}
\nonumber\\ &&\vdots\nonumber\\
\hspace{-0.3in}&&\int D \Omega_j(x_1)\,(2j+1)\sum^{j}_{m^{(1)}_j =
-j}\!\!\!  e^{\textstyle i\, g\, m^{(1)}_j \int_{C_{x_2 x_1}} 
\sqrt{g_{\mu\nu}[A^{\Omega}](x)\,dx_{\mu}dx_{\nu}}}.
\end{eqnarray}
The magnetic quantum number $m^{(k)}_j\,(k=1, \ldots, n)$ belongs to
the infinitesimal segment $C_{x_{k+1} x_k}$, where $C_{x_{n+1} x_n} =
C_{x_1 x_n}$.

In compact form Eq.(\ref{label2.21}) can be written as a path
integral over gauge functions
\begin{eqnarray}\label{label2.22}
\hspace{-0.2in}W_j(C)= \frac{1}{(2j+1)^2}\int \prod_{x\in C} D \Omega_j(x) \, \sum_{\left\{ m_j(x) \right\} }\!\!\!(2j+1)\,e^{\textstyle i g
\oint_{C} m_j(x)\,\sqrt{g_{\mu\nu}[A^{\Omega}](x)\,dx_{\mu}dx_{\nu}}}.
\end{eqnarray}
The integrals along the infinitesimal segments $C_{x_k x_{k-1}}$ we
determine as [2]
\begin{eqnarray}\label{label2.23}
\int_{C_{x_k x_{k-1}}}\!\!\!
m_j(x)\,\sqrt{g_{\mu\nu}[A^{\Omega}](x)\,dx_{\mu}dx_{\nu}}&=&
m_j(x_{k-1})\, \sqrt{g_{\mu\nu}[A^{\Omega}](x_{k-1})\,\Delta
x_{\mu}\,\Delta x_{\nu}}=\nonumber\\
&=& m^{(k-1)}_j\sqrt{g_{\mu\nu}[A^{\Omega}](x_{k-1})\,\Delta
x_{\mu}\,\Delta x_{\nu}}.
\end{eqnarray}
where $\Delta x = x_k - x_{k-1}$.

Comparing the path integral (\ref{label2.22}) with that
suggested in Eq.(23) of Ref.[3] one finds rather strong
disagreement. First, this concerns the contribution of different
states $m_j$ of the representation $j$. In the case of the path
integral (\ref{label2.22}) there is a summation over all values of
the magnetic colour quantum number $m_j$, whereas the representation
of Ref.[3] contains only one term with $m_j = j$. Second, Ref.[3] claims that in the integrand of their path integral the
exponent should depend only on the gauge field projected onto the
third axis in colour space. However, this is only possible
if the gauge functions are slowly varying with $x$,
i.e. $\Omega(x_k)\Omega^{\dagger}(x_{k-1})\simeq 1$.  In this case the
parallel transport operator $U^{\Omega}(C_{x_k x_{k-1}})$ would read
[13]
\begin{eqnarray}\label{label2.24}
U^{\Omega}(C_{x_k x_{k-1}})= \exp{i\,g\,\int_{C_{x_k x_{k-1}}} d
x_{\mu}\,A^{\Omega}_{\mu}(x)} = 1 + i\,g\,(x_i - x_{i-1})\cdot
A^{\Omega}(x_{i-1}),
\end{eqnarray}
and the evaluation of the character $\chi[U^{\Omega}_j(C_{x_k
x_{k-1}}]$ would run as follows
\begin{eqnarray}\label{label2.25}
\hspace{-0.3in}&&<m_j|[U^{\Omega}_j(C_{x_k x_{k-1}})]|m_j> =
 1 +(t^a_j)_{m_j m_j}\,
i\,g\,(x_k - x_{k-1})\cdot [A^{\Omega}(x_{k-1})]^{(a)}=\nonumber\\
\hspace{-0.3in}&&= 1 + m_j\,i\,g\,(x_k - x_{k-1})\cdot
[A^{\Omega}(x_{k-1})]^{(3)} = e^{\textstyle 
i\,g\int_{C_{x_k x_{k-1}}}dx_{\mu}\,m_j(x)[A^{\Omega}_{\mu}(x)]^{(3)}}\!\!\!,
\end{eqnarray}
where we have used the matrix elements of the generators of
$SU(2)$, i.e. $(t^a_j)_{m_j m_j} =
m_j\,\delta^{a3}$. More generally the exponent on the
r.h.s. of Eq.(\ref{label2.25}) can be written as
\begin{eqnarray}\label{label2.26}
\int_{C_{x_k x_{k-1}}}dx_{\mu}\,m_j(x)\,[A^{\Omega}_{\mu}(x)]^{(3)} =
2\int_{C_{x_k x_{k-1}}}dx_{\mu}\,m_j(x)\,{\rm tr}[t^3_j A^{\Omega}_{\mu}(x)].
\end{eqnarray}
This gives the path integral representation for Wilson loops
defined for $SU(2)$ in the following form
\begin{eqnarray}\label{label2.27}
\hspace{-0.3in}&&W_j(C)=\frac{1}{(2j+1)^2}\int \prod_{x\in C} D \Omega_j(x) \, \sum_{\left\{ m_j(x) \right\} }\!\!\!(2j+1)\,
e^{\textstyle 2 i  g \oint_{C}dx_{\mu}\,m_j(x)\,{\rm tr}[t^3_j A^{\Omega}_{\mu}(x)]}.\nonumber\\
\hspace{-0.3in}&&
\end{eqnarray}
The exponent contains the gauge field projected onto the third axis in
colour space ${\rm tr}[t^3_j A^{\Omega}_{\mu}(x)]$.
 Nevertheless, Eq.(\ref{label2.27}) differs form Eq.(23)
of Ref.[3] by a summation over all values of the colour magnetic quantum
number $m_j$ of the given irreducible representation $j$.

The repeated application of Eq.(\ref{label2.2}) induces that the
integrations over the gauge function at $x_k$ are completely
independent of the integrations at $x_{k\pm 1}$. There is no mechanism
which leads to gauge functions smoothly varying with $x_k\,(k =
1,\ldots\,n)$. In this sense the situation is opposite to the quantum
mechanical path integral. In Quantum Mechanics the integration over
all paths is restricted by the kinetic term of the Lagrange function.
In the semiclassical limit $\hbar \to 0$ due to the kinetic term the
fluctuations of all trajectories are shrunk to zero around a
classical trajectory. However, in the case of the integration over
gauge functions for the path integral representation of the Wilson
loop, there is neither a suppression factor nor a semiclassical limit
like $\hbar \to 0$. The key point of the application of
Eq.(\ref{label2.2}) and, therefore, the path integral representation
for Wilson loops is that all integrations over $\Omega(x_k)\,(k =
1,\ldots,n)$ are completely independent and can differ substantially
even if the points, where the gauge functions $\Omega(x_k)$ and
$\Omega(x_{k-1})$ are defined, are infinitesimally close to each other.

For the derivation of Eq.(23) of Ref.[3] Diakonov and Petrov have used
at an intermediate step a regularization drawing an analogy with an
axial--symmetric top with moments of inertia $I_{\perp}$ and
$I_{\parallel}$. Within this regularization the evolution operator
representing Wilson loops has been replaced by a path integral
over dynamical variables of this axial--symmetric top which correspond to gauge degrees of freedom of the non--Abelian gauge field.  The regularized expression of the evolution
operator has been obtained in the limit $I_{\perp}, I_{\parallel} \to
0$. The moments of inertia have been used as parameters
like $\hbar \to 0$. Unfortunately, as we show in Sect.\,6
the limit $I_{\perp}, I_{\parallel} \to 0$ has been evaluated
incorrectly.

\section{The $SU(N)$ extension}
\setcounter{equation}{0}

\hspace{0.2in} The extension of the path integral representation given in
Eq.(\ref{label2.24}) to $SU(N)$ is rather straightforward and reduces
to the evaluation of the character of the matrix $U^{\Omega}_r(C_{x_k
x_{k-1}})$ in the irreducible representation $r$ of $SU(N)$. The
character can be given by [12]
\begin{eqnarray}\label{label3.1}
\chi[U^{\Omega}_r(C_{x_k x_{k-1}})] &=& {\rm tr}(e^{\textstyle 
i\sum^{N-1}_{{\ell} = 1} H_{\ell}\Phi_{\ell}[C_{x_k x_{k-1}};
A^{\Omega}]}) = \nonumber\\
&=& \sum_{\vec{m}_r}\gamma_{\,\vec{m}_r}\,e^{\textstyle 
i\,\vec{m}_r\cdot {\vec{\Phi}}[C_{x_k x_{k-1}}; A^{\Omega}]},
\end{eqnarray}
where $H_{\ell}\,({\ell} = 1,\ldots,N-1)$ are diagonal $d_r\times d_r$
traceless matrices realizing the representation of the Cartan
subalgebra, i.e.  $[H_{\ell},H_{\ell'}]=0$, of the generators of the
$SU(N)$ [12].  The sum runs over all the weights $\vec{m}_r
=(m_{r\,1},\ldots,m_{r\,N-1})$ of the irreducible
representation $r$ and $\gamma_{\,\vec{m}_r}$ is the
multiplicity of the weight $\vec{m}_r$ and
$\sum_{\vec{m}_r}\gamma_{\,\vec{m}_r} = d_r$.  The
components of the vector ${\vec{\Phi}}[C_{x_k x_{k-1}}; A^{\Omega}]$
are defined by
\begin{eqnarray}\label{label3.2}
\Phi_{\ell}[C_{x_k x_{k-1}}; A^{\Omega}] = g\int_{C_{x_k
x_{k-1}}}\!\!\! {\varphi}_{\ell}\,[\omega(x)],
\end{eqnarray}
where we have introduced the notation $\omega(x) = t^a\omega^a(x) =
dz\cdot A^{\Omega}(x)$. The functions ${\varphi}_{\ell}\,[\omega(x)]$
are proportional to the roots of the equation ${\rm det}\Big[\omega(x)
- \lambda\Big]=0$.

The path integral representation of Wilson loops
defined for the irreducible representation $r$ of $SU(N)$ reads
\begin{eqnarray}\label{label3.3}
\hspace{-0.3in}W_r(C) =\frac{1}{d^2_r} 
\int \prod_{x\in C} D\Omega_r(x) \sum_{\left\{ \vec{m}_r(x) \right\} } \,d_r\,\gamma_{\,\vec{m}_r(x)}\, e^{\textstyle
i\,g\,\oint_{C}\vec{m}_r(x)\cdot \vec{\varphi}\,[\omega(x)]}.
\end{eqnarray}
Let us consider in more details the path integral representation of
Wilson loops defined for the fundamental representation
$\underline{3}$ of $SU(3)$. The character
$\chi_{\underline{3}}[U^{\Omega}_{\underline{3}}(C)]$ is defined as
\begin{eqnarray}\label{label3.4}
\hspace{-0.3in}&&\chi_{\underline{3}}[U^{\Omega}_{\underline{3}}(C)]
= {\rm tr}\Big(e^{\textstyle  iH_1\Phi_1[C; A^{\Omega}] + iH_2\Phi_2[C;
A^{\Omega}]}\,\Big) = e^{\textstyle - i\Phi_2[C; A^{\Omega}]/3} \nonumber\\
\hspace{-0.3in}&&+ e^{\textstyle i\Phi_1[C;
A^{\Omega}]/2\sqrt{3}}\,e^{\textstyle i\Phi_2[C;
A^{\Omega}]/6} + e^{\textstyle -
i\Phi_1[C; A^{\Omega}]/2\sqrt{3}}\,e^{\textstyle  i\Phi_2[C;
A^{\Omega}]/6},
\end{eqnarray}
where $H_1 = t^3/\sqrt{3}$ and $H_2 = t^8/\sqrt{3}$ [12]. For the
representation $\underline{3}$ of $SU(3)$ the equation
${\rm det}[\omega - \lambda]=0$ takes the form
\begin{eqnarray}\label{label3.5}
\lambda^3 - \lambda\,\frac{1}{2}\,{\rm tr}\,\omega^2(x) - {\rm
det}\,\omega(x) = 0.
\end{eqnarray}
The roots of Eq.(\ref{label3.5}) read
\begin{eqnarray}\label{label3.6}
\lambda^{(1)}&=& - \frac{1}{\sqrt{6}}\,\sqrt{{\rm
tr}\,\omega^2(x)}\,\cos\Bigg(\frac{1}{3}\,{\rm
arccos}\sqrt{2\,{\rm det}\Bigg[\displaystyle 1 +
12\,\frac{t^a{\rm tr}(t^a \omega^2(x))}{{\rm
tr}\,\omega^2(x)}\Bigg]}\,\Bigg) \nonumber\\ &&-
\frac{1}{\sqrt{2}}\,\sqrt{{\rm
tr}\,\omega^2(x)}\,\sin\Bigg(\frac{1}{3}\,{\rm
arccos}\sqrt{2\,{\rm det}\Bigg[\displaystyle 1 +
12\,\frac{t^a{\rm tr}(t^a \omega^2(x))}{{\rm
tr}\,\omega^2(x)}\Bigg]}\,\Bigg) ,\nonumber\\ \lambda^{(2)}&=& -
\frac{1}{\sqrt{6}}\,\sqrt{{\rm
tr}\,\omega^2(x)}\,\cos\Bigg(\frac{1}{3}\,{\rm
arccos}\sqrt{2\,{\rm det}\Bigg[\displaystyle 1 +
12\,\frac{t^a{\rm tr}(t^a \omega^2(x))}{{\rm
tr}\,\omega^2(x)}\Bigg]}\,\Bigg)  \nonumber\\ &&+
\frac{1}{\sqrt{2}}\,\sqrt{{\rm
tr}\,\omega^2(x)}\,\sin\Bigg(\frac{1}{3}\,{\rm
arccos}\sqrt{2\,{\rm det}\Bigg[\displaystyle 1 +
12\,\frac{t^a{\rm tr}(t^a \omega^2(x))}{{\rm
tr}\,\omega^2(x)}\Bigg]}\,\Bigg) ,\nonumber\\
\lambda^{(3)}&=&~\,\sqrt{\frac{2}{3}}\,\sqrt{{\rm
tr}\,\omega^2(x)}\,\cos\Bigg(\frac{1}{3}\,{\rm
arccos}\sqrt{2\,{\rm det}\Bigg[\displaystyle 1 +
12\,\frac{t^a{\rm tr}(t^a \omega^2(x))}{{\rm
tr}\,\omega^2(x)}\Bigg]}\,\Bigg) .\nonumber\\ &&
\end{eqnarray}
In terms of the roots $\lambda^{(i)}\,(i =1,2,3)$ the phases
$\Phi_{1,2}[C; A^{\Omega}]$ are defined as
\begin{eqnarray}\label{label3.7}
\hspace{-0.3in}&&\Phi_1[C;
A^{\Omega}]= -g\sqrt{6}\oint_C\sqrt{{\rm
tr}\,\omega^2(x)}\sin\Bigg(\frac{1}{3}\,{\rm
arccos}\sqrt{2\,{\rm det}\Bigg[\displaystyle 1 +
12\frac{t^a{\rm tr}(t^a \omega^2(x))}{{\rm
tr}\,\omega^2(x)}\Bigg]}\,\Bigg),\nonumber\\
\hspace{-0.3in}&&\Phi_2[C;
A^{\Omega}]= - g\sqrt{6}\oint_C\sqrt{{\rm
tr}\,\omega^2(x)}\cos\Bigg(\frac{1}{3}{\rm
arccos}\sqrt{2\,{\rm det}\Bigg[\displaystyle 1 +
12\frac{t^a{\rm tr}(t^a \omega^2(x))}{{\rm
tr}\,\omega^2(x)}\Bigg]}\,\Bigg),
\end{eqnarray}
where ${\rm tr}\,\omega^2(x) =
\frac{1}{2}\,g_{\mu\nu}[A^{\Omega}](x)\,dx_{\mu}dx_{\nu}$. Thus, in
the fundamental representation $\underline{3}$ the path integral representation for Wilson loops reads
\begin{eqnarray}\label{label3.8}
\hspace{-0.3in}W_{\underline{3}}(C)&=&\frac{1}{9}\int \prod_{x\in
C} \left[D \Omega_{\underline{3}}(x)\times\,3 \right] \, \Big(
e^{\textstyle i\Phi_1[C; A^{\Omega}]/2\sqrt{3}}e^{\textstyle
i\Phi_2[C; A^{\Omega}]/6} \nonumber\\
\hspace{-0.3in}&& + e^{\textstyle -
i\Phi_1[C; A^{\Omega}]/2\sqrt{3}}\,e^{\textstyle  i\Phi_2[C;
A^{\Omega}]/6} + e^{\textstyle - i\Phi_2[C; A^{\Omega}]/3}\Big),
\end{eqnarray}
where the phases $\Phi_{1,2}[C; A^{\Omega}]$ are given by
Eq.(\ref{label3.7}).

\section{Wilson loop for pure gauge field}
\setcounter{equation}{0}

\hspace{0.2in} As has been pointed out in Ref.[3] the path integral over gauge
degrees of freedom representing Wilson loops {\it is not of the
Feynman type, therefore, it depends explicitly on how one
``understands'' it, i.e. how it is discretized and regularized}. We
would like to emphasize that the {\it regularization procedure}
applied in Ref.[3] has led to an expression for Wilson loops which
supports the hypothesis of Maximal Abelian Projection [14]. According
to this hypothesis only Abelian degrees of freedom of non--Abelian
gauge fields are responsible for confinement. This is to full extent a
dynamical hypothesis. It is quite obvious that such a dynamical
hypothesis cannot be derived only by means of a regularization
procedure.

In order to show that the problem touched in this paper is not of
marginal interest and to check if path integral expressions that look differently superficially could actually compute the same number we evaluate below
explicitly the path integrals representing Wilson loop for a pure
$SU(2)$ gauge field. As has been stated in Ref.[3] for Wilson loops
$C$ a gauge field {\it along a given curve can be always written as a
``pure gauge''} and the derivation of the path integral representation
for Wilson loops can be provided for the gauge field taken {\it
without loss of generality in the ``pure gauge'' form}. We would like
to show that for the pure $SU(2)$ gauge field the path integral
representation for Wilson loops suggested in Ref.[3] fails for a
correct description of Wilson loops. Since a pure gauge field is
equivalent to a zero gauge field Wilson loops should be unity.

Of course, any correct path integral representation for Wilson
loops should lead to the same result. The evaluation of Wilson loops
within the path integral representation Eq.(\ref{label2.8}) is rather
trivial and transparent. Indeed, we have not corrupted the starting
expression for Wilson loops (\ref{label2.1}) by any artificial
regularization. Thereby, the general formula (\ref{label2.8})
evaluated through the discretization given by Eqs.(\ref{label2.7}) and
(\ref{label2.6}) is completely identical to the original expression
(\ref{label2.1}). The former gives a unit value for Wilson loops
defined for an arbitrary contour $C$ and an irreducible representation
$J$ of $SU(2)$: $W_J(C)=1$. 

Let us focus now on the path integral representation suggested in
Ref.[3]
\begin{eqnarray}\label{label4.1}
W_J(C)=\int \prod_{x\in C} D \Omega(x)\,e^{\textstyle 2 i J g 
\oint_{C}dx_{\mu}\,{\rm tr}[t^3 A^{\Omega}_{\mu}(x)]},
\end{eqnarray}
where all matrices are taken in the irreducible representation $J$.
Following the discretization suggested in Ref.[3] we arrive at the
expression
\begin{eqnarray}\label{label4.2}
W_J(C)= \lim_{n\to \infty}\prod^{n}_{k=1}\int D
\Omega(x_k)\,e^{\textstyle 2 i J g\int_{C_{x_{k+1} x_k}}dx_{\mu}\,{\rm
tr}[t^3 A^{\Omega}_{\mu}(x)]}.
\end{eqnarray}
Setting $A_{\mu}(x)=\partial_{\mu}U(x) U^{\dagger}(x)/ig$ we get
\begin{eqnarray}\label{label4.3}
A^{\Omega}_{\mu}(x)=\frac{1}{ig}\partial_{\mu}(\Omega(x)U(x))(\Omega(x)
U(x))^{\dagger}.
\end{eqnarray}
By a gauge transformation $\Omega(x)U(x) \to
\Omega(x)$ we reduce Eq.(\ref{label4.1}) to the form 
\begin{eqnarray}\label{label4.4}
W_J(C)=\int \prod_{x\in C} D \Omega(x)\,e^{\textstyle 2J
\oint_{C}dx_{\mu}\,{\rm tr}[t^3\partial_{\mu}\Omega(x)
\Omega^{\dagger}(x)]}.
\end{eqnarray}
For simplicity we consider Wilson loops in the fundamental
representation of $SU(2)$, $W_{1/2}(C)$. The result can be generalized
to any irreducible representation $J$.

For the evaluation of the path integral Eq.(\ref{label4.4}) it is
convenient to use a standard $s$--parameterization of Wilson loops
$C$ [2]: $x_{\mu} \to x_{\mu}(s)$, with $s \in
[0,1]$ and $x_{\mu}(0) = x_{\mu}(1) = x_{\mu}$.

The Wilson loop (\ref{label4.4}) reads in the $s$--parameterization
\begin{eqnarray}\label{label4.5}
W_{1/2}(C)=\int \prod_{0 \le s \le 1} D \Omega(s)\,\exp  \int\limits^1_0
ds\,{\rm tr}\Bigg[t^3\frac{d\Omega(s)}{ds}\Omega^{\dagger}(s)\Bigg].
\end{eqnarray}
The discretized form of the path integral (\ref{label4.5}) is given by 
\begin{eqnarray}\label{label4.6}
&&W_{1/2}(C) = \lim_{n \to\infty}\int \prod^n_{k = 1} D \Omega_k\,\exp 
\Delta s_{k+1,k}\,{\rm
tr}\Bigg[t^3\frac{\Omega_{k+1}-\Omega_{k}}{\Delta
s_{k+1,k}}\Omega^{\dagger}_k\Bigg]=\nonumber\\ 
&& = \lim_{n
\to\infty}\int \prod^n_{k = 1} D \Omega_k\,e^{\textstyle {\rm
tr}[t^3\Omega_{k+1}\Omega^{\dagger}_k]} = \lim_{n\to \infty}\int
\ldots \int D\Omega_n D\Omega_{n-1} D\Omega_{n-2}\ldots
D\Omega_1\,\nonumber\\ &&\times\, e^{\textstyle {\rm tr}[t^3\Omega_n
\Omega^{\dagger}_{n-1}]}\, e^{\textstyle 2J{\rm
tr}[t^3\Omega_{n-1}\Omega^{\dagger}_{n-2}]} \ldots e^{\textstyle  {\rm
tr}[t^3\Omega_2 \Omega^{\dagger}_1]} \, e^{\textstyle {\rm
tr}[t^3\Omega_1 \Omega^{\dagger}_n]},
\end{eqnarray}
where $\Omega_{n+1} = \Omega_{1}$.

For the subsequent integration we would use the expansion [15]
\begin{eqnarray}\label{label4.7}
e^{\textstyle z{\rm tr}[t^3\Omega]} = \sum_{j}
a_j(z)\,\chi_j\Big[e^{\textstyle i\pi t^3}\Omega\Big].
\end{eqnarray}
When the exponent of the l.h.s. is defined for the fundamental
representation, the coefficients $a_j(z)$ are given by [15,16]
\begin{eqnarray}\label{label4.8}
a_j(z) = e^{\textstyle -i\pi j}\,(2j+1)\,\frac{2 J_{2j+1}(z)}{z},
\end{eqnarray}
where $J_{2j+1}(z)$ are Bessel functions and the index $j$ runs over
$j =0,1/2,1,3/2,2,\ldots$ [16]. Recall, that $z = 2J$ and for the
fundamental representation $J = 1/2$ we should set $z = 1$.

For the integration over $\Omega_k$ we
suggest to use a formula of Ref.[17] modified for our case
\begin{eqnarray}\label{label4.9}
\int D \Omega\,e^{\textstyle z{\rm tr}[t^3 A\Omega^{\dagger} +
Bt^3\Omega]} = \sum_{j}
\frac{a^2_j(z)}{2j+1}\,\chi_j\Big[\Big(e^{\textstyle i\pi
t^3}\Big)^2AB\Big],
\end{eqnarray}
where the coefficients $a_j(z)$ are defined by the expansion
Eq.(\ref{label4.7}). The formula Eq.(\ref{label4.9}) can be derived by
using the orthogonality relation for characters [9,10,17]
\begin{eqnarray}\label{label4.10}
\int D\Omega \,\chi_j[A \Omega^{\dagger}]\,\chi_{j^{\prime}}[\Omega B] =
\frac{\delta_{j j^{\prime}}}{2j+1}\,\chi_j[AB].
\end{eqnarray}
Integrating over $\Omega_i\,(i = 1,2,\ldots,n)$ we arrive at the
expression
\begin{eqnarray}\label{label4.11}
W_{1/2}(C) &=& \lim_{n\to \infty}\sum_{j} (2j +
1)\,\Bigg[\frac{a_j}{2j+1}\Bigg]^n\,\chi_j\Big[\Big(e^{\textstyle i\pi
t^3}\Big)^n\Big]=\nonumber\\
&=&\lim_{n\to \infty}\sum_{j} (2j +
1)\,[2J_{2+1}(1)]^n\,\sum^{2j}_{k=0}e^{\textstyle -i\pi k n},
\end{eqnarray}
where we have denoted $a_j(1) = a_j = e^{\textstyle -i\pi
j}\,(2j+1)\,2 J_{2j+1}(1)$.  The series over $j$ is convergent for any
finite $n$ and every term of this series vanishes at $n\to
\infty$. This proves that the Wilson loop $W_{1/2}(C)$ vanishes in the
limit $n \to \infty$, i.e. $W_{1/2}(C) = 0$.

Thus, the Wilson loop $W_{1/2}(C)$ for an arbitrary contour $C$ and a
pure gauge field represented by the path integral derived in Ref.[3]
vanishes, instead of being equal to unity, $W_{1/2}(C) = 1$. This
shows that the path integral representation suggested in Ref.[3] fails
for the correct description of Wilson loops.

\section{Wilson loop for $Z(2)$ center vortices}
\setcounter{equation}{0}

\hspace{0.2in} In this Section we evaluate explicitly the path integral
 (\ref{label4.1}) for Wilson loops pierced by a $Z(2)$ center vortex
 with spatial azimuthal symmetry.  Some problems of $Z(2)$ center
 vortices with spatial azimuthal symmetry have been analysed by
 Diakonov in his recent publication [18] for the gauge group $SU(2)$.
 In this system the main dynamical variable is the azimuthal component
 of the non--Abelian gauge field $A^a_{\phi}(\rho)\,$ ($a = 1,2,3$)
 depending only on $\rho$, the radius in the transversal plane.  For a
 circular Wilson loop in the irreducible representation $J$ one gets
\begin{eqnarray}\label{label5.1}
W_J(\rho) = \frac{1}{2J+1}\sum^{J}_{m = - J}e^{\textstyle i 2\pi m 
\mu(\rho)}= \frac{1}{2J+1}\frac{\sin[(2J+1)\pi
\mu(\rho)]}{\sin[\pi \mu(\rho)]},
\end{eqnarray}
where $\mu(\rho) = \rho \sqrt{A^a_{\phi}(\rho) A^a_{\phi}(\rho)}$. The
gauge coupling constant $g$ is included in the definition of the gauge
field. For Wilson loops in the fundamental representation $J=1/2$
we have
\begin{eqnarray}\label{label5.2}
W_{1/2}(\rho) = \cos[\pi \mu(\rho)].
\end{eqnarray}
In the case of $Z(2)$ center vortices with spatial azimuthal symmetry
and for the fundamental representation of $SU(2)$ Eq.(\ref{label4.2}) takes the form
\begin{eqnarray}\label{label5.3}
W_{1/2}(\rho)&=& \lim_{n\to \infty}\prod^{n}_{k=1}\int D
\Omega_k\,e^{\textstyle {\rm tr}[t^3(i2\pi\rho/n)\Omega_{k+1}
A_{\phi}(\rho)\Omega^{\dagger}_k +
t^3 \Omega_{k+1}\Omega^{\dagger}_k]}= \nonumber\\ 
&&=\lim_{n\to \infty}\int \ldots 
\int D\Omega_n D\Omega_{n-1} D\Omega_{n-2}\ldots
D\Omega_1 \nonumber\\
&&\times\,e^{\textstyle 
\,{\rm tr}[t^3(i2\pi\rho/n)\Omega_n  A_{\phi}(\rho)\Omega^{\dagger}_{n-1}
 + t^3\Omega_n \Omega^{\dagger}_{n-1}]}\nonumber\\
&&\times\,e^{\textstyle {\rm tr}[t^3(i2\pi\rho/n)\Omega_{n-1}
 A_{\phi}(\rho)\Omega^{\dagger}_{n-2} + t^3\Omega_{n-1}
\Omega^{\dagger}_{n-2}]}\,\ldots \nonumber\\
&&\times\,e^{\textstyle 
\,{\rm tr}[t^3(i2\pi\rho/n)\Omega_2 A_{\phi}(\rho)\Omega^{\dagger}_1 + 
t^3\Omega_2 \Omega^{\dagger}_1]}\nonumber\\
&&\times\,e^{\textstyle {\rm tr}[t^3(i2\pi\rho/n)\Omega_1  
A_{\phi}(\rho)\Omega^{\dagger}_n + t^3\Omega_1\Omega^{\dagger}_n]},
\end{eqnarray}
where we have used $C_{x_{k+1} x_k} = 2\pi\rho/n$, $\Omega(x_k) = \Omega_k$ and $\Omega_{n+1} = \Omega_1$.

For the subsequent evaluation it is convenient to introduce the matrix
\begin{eqnarray}\label{label5.4}
Q(A_{\phi}) = \Big(1 + i\,\frac{2\pi}{n}\,\rho\, A_{\phi}(\rho)\Big).
\end{eqnarray}
In terms of $Q(A_{\phi})$ the path integral (\ref{label5.3}) reads
\begin{eqnarray}\label{label5.5}
&&W_{1/2}(\rho) = \lim_{n\to \infty}\int \ldots \int D\Omega_n
D\Omega_{n-1} D\Omega_{n-2}\ldots D\Omega_1\,e^{\textstyle {\rm
tr}[t^3\Omega_n Q(A_{\phi})\Omega^{\dagger}_{n-1}]}\, \nonumber\\
&&\times\,e^{\textstyle {\rm tr}[t^3\Omega_{n-1}Q(A_{\phi})
\Omega^{\dagger}_{n-2}]}\,\ldots \,e^{\textstyle {\rm tr}[ t^3\Omega_2
Q(A_{\phi})\Omega^{\dagger}_1]}\,e^{\textstyle \,{\rm tr}[t^3\Omega_1
Q(A_{\phi})\Omega^{\dagger}_n]},
\end{eqnarray}
The integration over $\Omega_k$ we carry out with the help of
Eq.(\ref{label4.9}) taken in the from
\begin{eqnarray}\label{label5.6}
&&\int D \Omega_k\,e^{\textstyle {\rm tr}[t^3\Omega_{k+1}
Q(A_{\phi})\Omega^{\dagger}_k + Q(A_{\phi})
\Omega^{\dagger}_{k-1}t^3\Omega_k]} = \nonumber\\
&&=\sum_{j}
\frac{a^2_j}{2j+1}\,\chi_j\Big[\Big(e^{\textstyle i\pi t^3}\Big)^2\Omega_{k+1}
Q^2(A_{\phi})\Omega^{\dagger}_{k-1}\Big].
\end{eqnarray}
and the orthogonality relation for the group characters. This yields
\begin{eqnarray}\label{label5.7}
W_{1/2}(\rho)= \lim_{n\to \infty} \sum_{j}
\Bigg[\frac{a_j}{2j+1}\Bigg]^n\, \chi_j\Big[\Big(e^{\textstyle i\pi
t^3}\Big)^n\Big]\,\chi_j[Q^n(A_{\phi})].
\end{eqnarray}
The evaluation of the character $\chi_j[Q^n(A_{\phi})]$ for $n \to
\infty$ runs as follows
\begin{eqnarray}\label{label5.8}
\chi_j[Q^n(A_{\phi})] &=& \chi_j\Big[\Big(1 +
i\,\frac{2\pi}{n}\,\rho\, A_{\phi}(\rho)\Big)^n\Big] =
\chi_j\Big[e^{\textstyle i 2\pi \rho A_{\phi}(\rho)}\Big] =
\nonumber\\ &=&\frac{\sin[(2j+1)\pi \mu(\rho)]}{\sin[\pi
\mu(\rho)]}.
\end{eqnarray}
Substituting Eq.(\ref{label5.8}) in Eq.(\ref{label5.7}) we obtain
\begin{eqnarray}\label{label5.9}
W_{1/2}(\rho)= \lim_{n\to \infty}\sum_{j}
[2J_{2j+1}(1)]^n\,\frac{\sin[(2j+1)\pi \mu(\rho)]}{\sin[\pi
\mu(\rho)]}\,\sum^{2j}_{k=0}e^{\textstyle -i\pi k n}
\end{eqnarray}
The series over $j$ is convergent for any finite $n$ and at $n \to
\infty$ every term vanishes. This gives $W_{1/2}(\rho) = 0$. Thus, we
have shown that the the path integral for Wilson loops suggested in
Ref.[3] gives zero for a field configuration with a $Z(2)$ center
vortex, $W_{1/2}(\rho) = 0$, instead of the correct result
$W_{1/2}(\rho) = \cos\pi\mu(\rho)$.

We hope that the examples considered in Sect.4 and 5 demonstrate that
the path integral representation for Wilson loops derived in Ref.[3]
is erroneous. Nevertheless, in Sect.\,6 we evaluate explicitly the
regularized evolution operator $Z_{\rm Reg}(R_2,R_1)$ suggested by
Diakonov and Petrov for the representation of the Wilson loop in
Ref.[3]. We show that this regularized evolution operator $Z_{\rm
Reg}(R_2,R_1)$ has been evaluated incorrectly in Ref.[3]. The correct
evaluation gives $Z_{\rm Reg}(R_2,R_1) = 0$ which agrees with our
results obtained above.

\section{Path integral for the evolution operator $Z(R_2,R_1)$ } 
\setcounter{equation}{0}

\hspace{0.2in} As has been suggested in Ref.[3] the functional  $Z(R_2,R_1)$
defined by (see Eq.(8) of Ref.[3])
\begin{eqnarray}\label{label6.1}
Z(R_2,R_1) =
\int\limits^{R_2}_{R_1}DR(t)\,\exp\Bigg(iT\int\limits^{t_2}_{t_1}\,{\rm
Tr}\,(iR\,\dot{R}\,\tau_3)\Bigg),
\end{eqnarray}
where $\dot{R} = dR/dt$ and $T = 1/2,1,3/2,\ldots$ is the colour
isospin quantum number, should be regularized by the analogy to an
axial--symmetric top. The regularized expression has been defined in Eq.(9) of Ref.[3] by
\begin{eqnarray}\label{label6.2}
Z_{\rm Reg}(R_2,R_1) =
\int\limits^{R_2}_{R_1}DR(t)\,\exp\Bigg(i\int\limits^{t_2}_{t_1}
\Big[\frac{1}{2}\,I_{\perp}\,(\Omega^2_1 + \Omega^2_2) +
\frac{1}{2}\,I_{\parallel}\,\Omega^2_3 + T\,\Omega_3\Big]\Bigg),
\end{eqnarray}
where $\Omega_a = i\,{\rm Tr}(R\,\dot{R}\,\tau_a)$ are angular
velocities of the top, $\tau_a$ are Pauli matrices $a=1,2,3$,
$I_{\perp}$ and $I_{\parallel}$ are the moments of inertia of the top
which should be taken to zero. According to the prescription of
Ref.[3] one should take first the limit $I_{\parallel} \to 0$ and
then $I_{\perp} \to 0$.

For the confirmation of the result, given in Eq.(13) of Ref.[3],  
\begin{eqnarray}\label{label6.3}
Z_{\rm Reg}(R_2,R_1) = (2T + 1)\,D^T_{TT}(R_2R^{\dagger}_1),
\end{eqnarray}
where $D^T(U)$ is a Wigner rotational matrix in the representation $T$,
the authors of Ref.[3] suggested to evaluate the evolution operator
(\ref{label6.2}) explicitly by means of the discretization of the
path integral over $R$.  The discretized form of the evolution
operator Eq.(\ref{label6.2}) is given by Eq.(14) of Ref.[3] and
reads\footnote{We are using the notations of Ref.[3]}
\begin{eqnarray}\label{label6.4}
\hspace{-0.5in}&&Z_{\rm Reg}(R_{N+1},R_0) = \lim_{\begin{array}{c} N\to
\infty\\ \delta \to 0\end{array}}{\cal N}\int\prod^{N}_{n=1}dR_n\nonumber\\ 
\hspace{-0.5in}&&\times\,\exp\Bigg[\sum^{N}_{n=0}
\Bigg(-\,i\,\frac{I_{\perp}}{2\delta}\,\Big[({\rm Tr}\,V_n\tau_1)^2 +
({\rm Tr}\,V_n\tau_2)^2\Big] - \,i\,\frac{I_{\parallel}}{2\delta}\,({\rm
Tr}\,V_n\tau_3)^2 -\,T\,({\rm Tr}\,V_n\tau_3)\Bigg)\Bigg],
\end{eqnarray}
where $R_n = R(s_n)$ with $s_n = t_1 + n\,\delta$ and $\Omega_a = i{\rm Tr}\,(R_n
R^{\dagger}_{n+1}\tau_a)/\delta$ is the discretized analogy of the
angular velocities [3] and $V_n = R_n R^{\dagger}_{n+1}$ are the
relative orientations of the top at neighbouring points. The
normalization factor ${\cal N}$ is determined by
\begin{eqnarray}\label{label6.5}
{\cal N} = \Bigg(\frac{I_{\perp}}{2\pi i
\delta}\sqrt{\frac{I_{\parallel}}{2\pi i \delta}}\,\Bigg)^{N+1}.
\end{eqnarray}
(see Eq.(19) of Ref.[3]). According to the prescription of Ref.[3] one
should take the limits $\delta \to 0$ and $I_{\parallel}, I_{\perp}
\to 0$ but keeping the ratios $I_i/\delta$, where ($ i= {\parallel},
{\perp}$), much greater than unity, $I_i/\delta \gg 1$.

Let us rewrite the exponent of the integrand of Eq.(\ref{label6.4}) in
equivalent form
\begin{eqnarray}\label{label6.6}
\hspace{-0.5in}&&Z_{\rm Reg}(R_{N+1},R_0) = \lim_{\begin{array}{c} N\to
\infty\\ \delta \to 0\end{array}}\Bigg(\frac{I_{\perp}}{2\pi i
\delta}\sqrt{\frac{I_{\parallel}}{2\pi i \delta}}\,\Bigg)^{N+1}\int\prod^{N}_{n=1}dR_n\nonumber\\ 
\hspace{-0.5in}&&\times\,\exp\Bigg[\sum^{N}_{n=0}
\Bigg(-\,i\,\frac{I_{\perp}}{2\delta}\,({\rm Tr}\,V_n\tau_a)^2 - 
\,i\,\frac{I_{\parallel}-I_{\perp}}{2\delta}\,({\rm
Tr}\,V_n\tau_3)^2 -\,T\,({\rm Tr}\,V_n\tau_3)\Bigg)\Bigg].
\end{eqnarray}
Now let us show that if $V_n$ is a rotation in the fundamental
representation of $SU(2)$, so 
\begin{eqnarray}\label{label6.7}
({\rm Tr}\,V_n\tau_a)^2
= -4 + ({\rm Tr}\,V_n)^2.
\end{eqnarray}
For this aim, first, recall that
\begin{eqnarray}\label{label6.8}
{\rm Tr}\,(V_n\tau_a) = - {\rm
Tr}\,(V^{\dagger}_n\tau_a).
\end{eqnarray}
Since $V_n$ is a rotation matrix in the fundamental representation of
$SU(2)$ [19]. By virtue of the
relation (\ref{label6.8}) we can rewrite $({\rm Tr}\,V_n\tau_a)^2$
as follows
\begin{eqnarray}\label{label6.9}
&&({\rm Tr}\,V_n\tau_a)^2 =  - {\rm Tr}\,(V_n\tau_a){\rm
Tr}\,(V^{\dagger}_n\tau_a) = - 2\, {\rm Tr}\,\Big(\Big(V_n -
\frac{1}{2}\,{\rm Tr}\,V_n\Big)\Big(V^{\dagger}_n- \frac{1}{2}\,{\rm
Tr}\,V_n\Big) \Big) =\nonumber\\ 
&&= - 2\, {\rm Tr}\,(R_n R^{\dagger}_{n+1} R_{n+1}R^{\dagger}_n) 
+ ({\rm Tr}\,V_n)^2 = - 2\,{\rm Tr}\,1+ ({\rm Tr}\,V_n)^2 
= - 4 + ({\rm Tr}\,V_n)^2.
\end{eqnarray}
By using the relation Eq.(\ref{label6.7}) we can recast the r.h.s. of
Eq.(\ref{label6.6}) into the form
\begin{eqnarray}\label{label6.10}
\hspace{-0.5in}&&Z_{\rm Reg}(R_{N+1},R_0) = \lim_{\begin{array}{c} N\to
\infty\\ \delta \to 0\end{array}} \Bigg[\Bigg(\frac{I_{\perp}}{2\pi i
\delta}\sqrt{\frac{I_{\parallel}}{2\pi i \delta}}\,\Bigg)^{N+1}\,
\exp\Bigg(i\,2\,(N+1)\,\frac{I_{\perp}}{\delta}\Bigg)\Bigg]
\nonumber\\ 
\hspace{-0.5in}&&\times\,\int\prod^{N}_{n=1}dR_n\,
\exp\Bigg[\sum^{N}_{n=0} \Bigg(-\,i\,\frac{I_{\perp}}{2\delta}\,({\rm
Tr}\,V_n)^2 -\,i\,\frac{I_{\parallel}-I_{\perp}}{2\delta}\,({\rm
Tr}\,V_n\tau_3)^2 -\,T\,({\rm Tr}\,V_n\tau_3)\Bigg)\Bigg].
\end{eqnarray}
Now let us proceed to the evaluation of the integrals over
$R_n\,(n=1,2,...,N)$. For this aim it is convenient to rewrite the
r.h.s. of Eq.(\ref{label6.10}) in the following form
\begin{eqnarray}\label{label6.11}
\hspace{-0.5in}&&Z_{\rm Reg}(R_{N+1},R_0) = \lim_{\begin{array}{c} N\to
\infty\\ \delta \to 0\end{array}} \Bigg[\Bigg(\frac{I_{\perp}}{2\pi i
\delta}\sqrt{\frac{I_{\parallel}}{2\pi i \delta}}\,\Bigg)^{N+1}\,
\exp\Bigg(i \,2\, (N+1)\,\frac{I_{\perp}}{\delta}\Bigg)\Bigg]
\nonumber\\ 
\hspace{-0.5in}&&\times\,\int\!\!\!\int\ldots \int\!\!\!\int  
dR_N\,dR_{N-1}\,\ldots\,dR_2\,dR_1\nonumber\\
\hspace{-0.5in}&&\times\,\exp\Bigg(-\,i\,\frac{I_{\perp}}{2\delta}\,
\Big[({\rm Tr}\,R_N R^{\dagger}_{N+1} )^2 + ({\rm
Tr}\,R_{N-1} R^{\dagger}_{N} )^2 + \ldots + ({\rm
Tr}\,R_2 R^{\dagger}_1 )^2 + ({\rm
Tr}\,R_1 R^{\dagger}_0 )^2\Big] \nonumber\\
\hspace{-0.5in}&&-\,i\,\frac{I_{\parallel}-I_{\perp}}{2\delta}\,\Big[({\rm
Tr}\,R_N R^{\dagger}_{N+1}\tau_3 )^2 + ({\rm
Tr}\,R_{N-1} R^{\dagger}_{N}\tau_3 )^2 + \ldots + ({\rm
Tr}\,R_2 R^{\dagger}_1\tau_3 )^2 + ({\rm
Tr}\,R_1 R^{\dagger}_0\tau_3 )^2\Big]\nonumber\\
\hspace{-0.5in}&& -T\,\Big[{\rm
Tr}\,(R_N R^{\dagger}_{N+1}\tau_3 ) + {\rm
Tr}\,(R_{N-1} R^{\dagger}_{N}\tau_3 ) + \ldots + {\rm
Tr}\,(R_2 R^{\dagger}_1\tau_3 ) + {\rm
Tr}\,(R_1 R^{\dagger}_0\tau_3 )\Big]\Bigg).
\end{eqnarray}
In the fundamental representation and the parameterization [19] we
have
\begin{eqnarray}\label{label6.12}
&&{\rm Tr}\,V_n ={\rm Tr}\,(R_n R^{\dagger}_{n+1}) =\nonumber\\
&&= 2\,\cos\frac{\beta_n}{2}\,\cos\frac{\beta_{n+1}}{2}\,
\cos\Bigg(\frac{\alpha_n + \gamma_n}{2} - \frac{\alpha_{n+1} +
\gamma_{n+1}}{2}\Bigg) \nonumber\\ && +
2\,\sin\frac{\beta_n}{2}\,\sin\frac{\beta_{n+1}}{2}\,
\cos\Bigg(\frac{\alpha_n - \gamma_n}{2} - \frac{\alpha_{n+1} -
\gamma_{n+1}}{2}\Bigg)=\nonumber\\ &&= 2\,\cos\Bigg(\frac{\beta_n
-\beta_{n+1} }{2}\Bigg)\, \cos\Bigg(\frac{\alpha_n -
\alpha_{n+1}}{2}\Bigg)\,\cos\Bigg(\frac{\gamma_n -
\gamma_{n+1}}{2}\Bigg) \nonumber\\ &&-
2\,\cos\Bigg(\frac{\beta_n + \beta_{n+1}}{2}\Bigg)\,
\sin\Bigg(\frac{\alpha_n - \alpha_{n+1}}{2}\Bigg)\,
\sin\Bigg(\frac{\gamma_n - \gamma_{n+1}}{2}\Bigg),\nonumber\\
&&{\rm Tr}\,(V_n\tau_3) = {\rm
Tr}\,(R_n R^{\dagger}_{n+1}\tau_3)=\nonumber\\
&&= -
2\,i\,\cos\frac{\beta_n}{2}\,\cos\frac{\beta_{n+1}}{2}\,
\sin\Bigg(\frac{\alpha_n + \gamma_n}{2} - \frac{\alpha_{n+1} +
\gamma_{n+1}}{2}\Bigg) \nonumber\\
&&+ 2\,i\,\sin\frac{\beta_n}{2}\,\sin\frac{\beta_{n+1}}{2}\,
\sin\Bigg(\frac{\alpha_n - \gamma_n}{2} - \frac{\alpha_{n+1} -
\gamma_{n+1}}{2}\Bigg)=\nonumber\\
&&=-\, 2\,i\cos\Bigg(\frac{\beta_n
-\beta_{n+1} }{2}\Bigg)\, \cos\Bigg(\frac{\alpha_n -
\alpha_{n+1}}{2}\Bigg)\,\sin\Bigg(\frac{\gamma_n -
\gamma_{n+1}}{2}\Bigg) \nonumber\\ 
&&- 2\,i\,\cos\Bigg(\frac{\beta_n + \beta_{n+1}}{2}\Bigg)\,
\sin\Bigg(\frac{\alpha_n - \alpha_{n+1}}{2}\Bigg)\,
\cos\Bigg(\frac{\gamma_n - \gamma_{n+1}}{2}\Bigg).
\end{eqnarray}
The Haar measure $R_n$ is defined by 
\begin{eqnarray}\label{label6.13}
DR_n = \frac{1}{8\pi^2}\,\sin\,\beta_n\,d\beta_n\,d\alpha_n\,d\gamma_n.
\end{eqnarray}
Due to the assumption $I_i/\delta \gg 1$, where ($ i= {\parallel},
{\perp}$), the integrals over $R_n$ are concentrated around unit
elements.  Expanding ${\rm Tr}\,(V_n)$ and ${\rm Tr}\,(V_n\tau_3)$
around unit elements we get
\begin{eqnarray}\label{label6.14}
\hspace{-0.5in}{\rm Tr}\,V_n &=& {\rm Tr}\,(R_n R^{\dagger}_{n+1}) =2
- \frac{1}{4}\,(\beta_n - \beta_{n+1})^2 - \frac{1}{4}\,(\alpha_n -
\alpha_{n+1} + \gamma_n - \gamma_{n+1})^2,\nonumber\\ \hspace{-0.5in}{\rm
Tr}\,(V_n\tau_3) &=& {\rm Tr}\,(R_n
R^{\dagger}_{n+1}\tau_3) = - \,i\,(\alpha_n -
\alpha_{n+1} + \gamma_n - \gamma_{n+1}).
\end{eqnarray}
For the subsequent integration it is convenient to make a change of variables
\begin{eqnarray}\label{label6.15}
\alpha_n + \gamma_n &\to& \gamma_n,\nonumber\\ \frac{\alpha_n -
\gamma_n}{2} &\to& \alpha_n.
\end{eqnarray}
The Jacobian of this transformation is equal to unity. After this
change of variables (\ref{label6.14}) reads
\begin{eqnarray}\label{label6.16}
\hspace{-0.5in}{\rm Tr}\,V_n &=& {\rm Tr}\,(R_n R^{\dagger}_{n+1}) =2
- \frac{1}{4}\,(\beta_n - \beta_{n+1})^2 - \frac{1}{4}\,(\gamma_n -
\gamma_{n+1})^2,\nonumber\\ \hspace{-0.5in}{\rm Tr}\,(V_n\tau_3) &=&
{\rm Tr}\,(R_n R^{\dagger}_{n+1}\tau_3) = - \,i\,(\gamma_n -
\gamma_{n+1}).
\end{eqnarray}
Since both ${\rm Tr}\,V_n$ and ${\rm Tr}\,(V_n\tau_3)$ do not depend
on $\alpha_n$, we can integrate out $\alpha_n$. This changes only the
Haar measure as follows
\begin{eqnarray}\label{label6.17}
DR_n = \frac{1}{8\pi}\,\beta_n\,d\beta_n\,d\gamma_n.
\end{eqnarray}
The integration over $\beta_n$ and $\gamma_n$ we will carry out in the
limits $-\infty \le \beta_n \le \infty$ and $-\infty \le \gamma_n \le
\infty$.

Substituting expansions (\ref{label6.16}) in the integrand of
Eq.(\ref{label6.11}) we obtain
\begin{eqnarray}\label{label6.18}
\hspace{-0.5in}&&Z_{\rm Reg}(R_{N+1},R_0) = \lim_{\begin{array}{c}
N\to \infty\\ \delta \to 0\end{array}} \Bigg(\frac{I_{\perp}}{2\pi i
\delta}\sqrt{\frac{I_{\parallel}}{2\pi i
\delta}}\,\Bigg)^{N+1}\,\Bigg(\frac{1}{8\pi}\Bigg)^N\nonumber\\
\hspace{-0.5in}&&\times
\int\limits^{\infty}_{-\infty}d\gamma_N
\int\limits^{\infty}_{-\infty}d\beta_N\,\beta_N
\int\limits^{\infty}_{-\infty}d\gamma_{N-1}
\int\limits^{\infty}_{-\infty}d\beta_{N-1}\,\beta_{N-1}\,\ldots
\int\limits^{\infty}_{-\infty}d\gamma_2
\int\limits^{\infty}_{-\infty}d\beta_2\,\beta_2
\int\limits^{\infty}_{-\infty}d\gamma_1
\int\limits^{\infty}_{-\infty}d\beta_1\,\beta_1
\nonumber\\
\hspace{-0.5in}&&\times\,\exp\Bigg(i\,\frac{I_{\perp}}{2\delta}\,
[(\beta_{N+1} - \beta_N)^2 + (\beta_N - \beta_{N-1})^2 + \ldots
+(\beta_2 - \beta_1)^2 + (\beta_1 - \beta_0)^2] \nonumber\\
\hspace{-0.5in}&&+\,i\,\frac{I_{\parallel}}{2\delta}\,[(\gamma_{N+1} -
\gamma_N)^2 + (\gamma_N - \gamma_{N-1})^2 + \ldots +(\gamma_2 -
\gamma_1)^2 + (\gamma_1 - \gamma_0)^2]\nonumber\\
\hspace{-0.5in}&& - \,i\,T\,[(\gamma_{N+1} - \gamma_N) + (\gamma_N
- \gamma_{N-1}) + \ldots +(\gamma_2 - \gamma_1) + (\gamma_1 -
\gamma_0)]\Bigg)=\nonumber\\
\hspace{-0.5in}&&= e^{\textstyle - \,i\,T\,(\gamma_{N+1} -
\gamma_0)}\lim_{\begin{array}{c} N\to \infty\\ \delta \to
0\end{array}} \Bigg(\frac{I_{\perp}}{2\pi i
\delta}\sqrt{\frac{I_{\parallel}}{2\pi i
\delta}}\,\Bigg)^{N+1}\,\Bigg(\frac{1}{8\pi}\Bigg)^N\nonumber\\
\hspace{-0.5in}&&\times
\int\limits^{\infty}_{-\infty}d\gamma_N
\int\limits^{\infty}_{-\infty}d\beta_N\,\beta_N
\int\limits^{\infty}_{-\infty}d\gamma_{N-1}
\int\limits^{\infty}_{-\infty}d\beta_{N-1}\,\beta_{N-1}\,\ldots
\int\limits^{\infty}_{-\infty}d\gamma_2
\int\limits^{\infty}_{-\infty}d\beta_2\,\beta_2
\int\limits^{\infty}_{-\infty}d\gamma_1
\int\limits^{\infty}_{-\infty}d\beta_1\,\beta_1
\nonumber\\
\hspace{-0.5in}&&\times\,\exp\Bigg(i\,\frac{I_{\perp}}{2\delta}\,
[(\beta_{N+1} - \beta_{N})^2 + (\beta_{N} - \beta_{N-1})^2 + \ldots
+(\beta_2 - \beta_1)^2 + (\beta_1 - \beta_0)^2] \nonumber\\
\hspace{-0.5in}&&+\,i\,\frac{I_{\parallel}}{2\delta}\,[(\gamma_{N+1} -
\gamma_{N})^2 + (\gamma_{N} - \gamma_{N-1})^2 + \ldots +(\gamma_2 -
\gamma_1)^2 + (\gamma_1 - \gamma_0)^2]\Bigg).
\end{eqnarray}
The integration over $\gamma_n$ gives
\begin{eqnarray}\label{label6.19}
\hspace{-0.3in}&&\int\limits^{\infty}_{-\infty}d\gamma_N
\int\limits^{\infty}_{-\infty}d\gamma_{N-1}
\,\ldots
\int\limits^{\infty}_{-\infty}d\gamma_2
\int\limits^{\infty}_{-\infty}d\gamma_1
\nonumber\\
\hspace{-0.3in}&&\times\,
\exp\Bigg(i\,\frac{I_{\parallel}}{2\delta}\,[(\gamma_{N+1} -
\gamma_{N})^2 + (\gamma_{N} - \gamma_{N-1})^2 + \ldots +(\gamma_2 -
\gamma_1)^2 + (\gamma_1 - \gamma_0)^2]\Bigg)=\nonumber\\
\hspace{-0.3in}&&=\sqrt{\frac{2\pi
i\delta}{I_{\parallel}}\,\frac{1}{2}}\sqrt{\frac{2\pi
i\delta}{I_{\parallel}}\,\frac{2}{3}} \ldots \sqrt{\frac{2\pi
i\delta}{I_{\parallel}}\,\frac{N-1}{N}}\sqrt{\frac{2\pi
i\delta}{I_{\parallel}}\,\frac{N}{N+1}}\,
\exp\Bigg(i\,\frac{I_{\parallel}}{2(N+1)\delta}\,(\gamma_{N+1} -
\gamma_0)^2\Bigg)=\nonumber\\
\hspace{-0.3in}&&= \Bigg(\sqrt{\frac{2\pi
i\delta}{I_{\parallel}}}\,\Bigg)^N \sqrt{\frac{1}{N+1}}\,\,
\exp\Bigg(i\,\frac{I_{\parallel}}{2(N+1)\delta}\,(\gamma_{N+1} -
\gamma_0)^2\Bigg).
\end{eqnarray}
By taking into account the normalization factor the result of the
integration over $\gamma_n$ reads
\begin{eqnarray}\label{label6.20}
\hspace{-0.5in}&&\Bigg(\sqrt{\frac{I_{\parallel}}{2\pi i
\delta}}\,\Bigg)^{N+1}\,\int\limits^{\infty}_{-\infty}d\gamma_N
\int\limits^{\infty}_{-\infty}d\gamma_{N-1}
\,\ldots
\int\limits^{\infty}_{-\infty}d\gamma_2
\int\limits^{\infty}_{-\infty}d\gamma_1
\nonumber\\
\hspace{-0.5in}&&\times\,
\exp\Bigg(i\,\frac{I_{\parallel}}{2\delta}\,[(\gamma_{N+1} -
\gamma_{N})^2 + (\gamma_{N} - \gamma_{N-1})^2 + \ldots +(\gamma_2 -
\gamma_1)^2 + (\gamma_1 - \gamma_0)^2]\Bigg)=\nonumber\\
\hspace{-0.5in}&&=\sqrt{\frac{I_{\parallel}}{2\pi i(N+1)\delta}}\,
\exp\Bigg(i\,\frac{I_{\parallel}}{2(N+1)\delta}\,(\gamma_{N+1} -
\gamma_0)^2\Bigg) = \nonumber\\
\hspace{-0.5in}&&=\sqrt{\frac{I_{\parallel}}{2\pi
i\Delta t}}\,
\exp\Bigg(i\,\frac{I_{\parallel}}{2\Delta t}\,(\gamma_{N+1} -
\gamma_0)^2\Bigg),
\end{eqnarray}
where we have replaced $(N+1)\,\delta = t_2 - t_1 = \Delta t$. The
obtained result is exact. By replacing $I_{\parallel} \to
M$, $\gamma_{N+1} \to x_b$, $\gamma_0 \to x_a$ and $\Delta t \to (t_b
- t_a)$ we arrive at the expression for the Green function, the
evolution operator, of a free particle with a mass $M$ given by
Eq.(2.51) of Ref.[6].

Thus, after the integration over $\gamma_n$ the evolution operator
$Z_{\rm Reg}(R_{N+1},R_0)$ can be written in the form
\begin{eqnarray}\label{label6.21}
\hspace{-0.5in}&&Z_{\rm Reg}(R_{N+1},R_0) =\nonumber\\
\hspace{-0.5in}&&=\sqrt{\frac{I_{\parallel}}{2\pi i\Delta t}}\,
\exp\Bigg(i\,\frac{I_{\parallel}}{2\Delta t}\,(\gamma_{N+1} -
\gamma_0)^2\Bigg)\,e^{\textstyle - \,i\,T\,(\gamma_{N+1} -
\gamma_0)}\,F[I_{\perp},\beta_{N+1},\beta_0],
\end{eqnarray}
where $F[I_{\perp},\beta_{N+1},\beta_0]$ is a functional defined by
the integrals over $\beta_n$
\begin{eqnarray}\label{label6.22}
\hspace{-0.5in}&&F[I_{\perp},\beta_{N+1},\beta_0] = \nonumber\\
\hspace{-0.5in}&&=\lim_{\begin{array}{c} N\to \infty\\ \delta \to
0\end{array}} \Bigg(\frac{I_{\perp}}{2\pi i
\delta}\Bigg)^{N+1}\,\Bigg(\frac{1}{4\pi}\Bigg)^N
\int\limits^{\infty}_{-\infty}d\beta_N\,\beta_N
\int\limits^{\infty}_{-\infty}d\beta_{N-1}\,\beta_{N-1}\,\ldots
\int\limits^{\infty}_{-\infty}d\beta_2\,\beta_2
\int\limits^{\infty}_{-\infty}d\beta_1\,\beta_1 \nonumber\\
\hspace{-0.5in}&&\times\,\exp\Bigg(i\,\frac{I_{\perp}}{2\delta}\,
[(\beta_{N+1} - \beta_{N})^2 + (\beta_{N} - \beta_{N-1})^2 + \ldots
+(\beta_2 - \beta_1)^2 + (\beta_1 - \beta_0)^2]\Bigg).
\end{eqnarray}
Formally we do not need to evaluate the functional
$F[I_{\perp},\beta_{N+1},\beta_0]$ explicitly. In fact, the functional
$F[I_{\perp},\beta_{N+1},\beta_0]$ should be a regular function of
variables $I_{\perp}$, $\beta_{N+1}$ and $\beta_0$ whose absolute
value is bound in the limit $I_{\perp} \to 0$. Therefore, taking the
limit $I_{\parallel} \to 0$ for the evolution operator $Z_{\rm
Reg}(R_{N+1},R_0)$ defined by Eq.(\ref{label6.21}) we get
\begin{eqnarray}\label{label6.23}
Z(R_2,R_1) = \lim_{I_{\parallel},I_{\perp} \to 0}Z_{\rm Reg}(R_{N+1},R_0) = 0.
\end{eqnarray}
This agrees with our results obtained in Sects.\,4 and 5.

Nevertheless, in spite of this very definite result let us proceed to
the explicit evaluation of the functional
$F[I_{\perp},\beta_{N+1},\beta_0]$ and show that the functional
$F[I_{\perp},\beta_{N+1},\beta_0]$ vanishes at $N \to \infty$. It is
convenient to rewrite the integrand of Eq.(\ref{label6.22}) in the
equivalent form
\begin{eqnarray}\label{label6.24}
\hspace{-0.5in}&&F[I_{\perp},\beta_{N+1},\beta_0]
=\lim_{\begin{array}{c} N\to \infty\\ \delta \to 0\end{array}}
\Bigg(\frac{I_{\perp}}{2\pi i
\delta}\Bigg)\,\Bigg(\frac{-1}{2\pi}\Bigg)^N\,\Bigg(\frac{1}{8\pi}\Bigg)^N
\nonumber\\
\hspace{-0.5in}&& \times\,
\frac{\partial}{\partial j_1}\frac{\partial}{\partial j_2}\ldots 
\frac{\partial}{\partial j_{N-1}}\frac{\partial}{\partial j_N}
\int\limits^{\infty}_{-\infty}d\beta_N
\int\limits^{\infty}_{-\infty}d\beta_{N-1}\ldots
\int\limits^{\infty}_{-\infty}d\beta_2
\int\limits^{\infty}_{-\infty}d\beta_1\nonumber\\
\hspace{-0.5in}&&\times\,\exp\Bigg(i\,\frac{I_{\perp}}{2\delta}\,
[(\beta_{N+1} - \beta_{N})^2 + (\beta_{N} - \beta_{N-1})^2 + \ldots
+(\beta_2 - \beta_1)^2 + (\beta_1 - \beta_0)^2\nonumber\\
\hspace{-0.5in}&& + 2\,j_N\,\beta_N +2\,j_{N-1}\,\beta_{N-1} + \ldots
+ 2\,j_2\, \beta_2 + 2\,j_1\,\beta_1 ]\Bigg)\Bigg|_{j_N = j_{N-1} =
\ldots =j_2 = j_1 = 0} .
\end{eqnarray}
After $k$ integrations we get
\begin{eqnarray}\label{label6.25}
\hspace{-0.5in}&&\int\limits^{\infty}_{-\infty}d\beta_k
\int\limits^{\infty}_{-\infty}d\beta_{k-1}\ldots
\int\limits^{\infty}_{-\infty}d\beta_2
\int\limits^{\infty}_{-\infty}d\beta_1\,
\exp\Bigg(i\,\frac{I_{\perp}}{2\delta}\,
[(\beta_{k+1} - \beta_{k})^2 + (\beta_{k} - \beta_{k-1})^2\nonumber\\
\hspace{-0.5in}&& + \ldots +(\beta_2 - \beta_1)^2 + (\beta_1 -
\beta_0)^2 + 2\,j_k\,\beta_k + 2\,j_{k-1}\,\beta_{k-1} + \ldots + 2\,j_2\,
\beta_2 + 2\,j_1\,\beta_1 ]\Bigg) =\nonumber\\
\hspace{-0.5in}&&=\sqrt{\frac{2\pi
i\delta}{I_{\perp}}\,\frac{1}{2}}\sqrt{\frac{2\pi
i\delta}{I_{\perp}}\,\frac{2}{3}} \ldots \sqrt{\frac{2\pi
i\delta}{I_{\perp}}\,\frac{k-1}{k}}\sqrt{\frac{2\pi
i\delta}{I_{\perp}}\,\frac{k}{k+1}}\,
\exp\Bigg(i\,\frac{I_{\perp}}{2(k+1)\delta}\,(
\beta_0 - \beta_{k+1})^2\Bigg)\nonumber\\
\hspace{-0.5in}&& 
\times\,\exp\Bigg(i\,\frac{I_{\perp}}{\delta}\,\beta_{k+1}\,
\Bigg(\frac{k}{k+1}\,j_k +
\frac{k}{k+1}\cdot\frac{k-1}{k}\,j_{k-1}
+\frac{k}{k+1}\cdot\frac{k-1}{k}\cdot\frac{k-2}{k-1}\,j_{k-2}\nonumber\\
\hspace{-0.5in}&& + \ldots +
\frac{k}{k+1}\cdot\frac{k-1}{k}\cdot\frac{k-2}{k-1}\cdots
\frac{2}{3}\,j_2 +
\frac{k}{k+1}\cdot\frac{k-1}{k}\cdot\frac{k-2}{k-1}\cdots
\frac{2}{3}\cdot\frac{1}{2}\,j_1\Bigg)\Bigg)\nonumber\\
\hspace{-0.5in}&&\times\,\exp\Bigg(i\,\frac{I_{\perp}}{\delta}\,
\Bigg[-\frac{1}{2}\cdot\frac{1}{2}\,\Big(j_1 - \beta_0\Big)^2 +
\frac{1}{2}\cdot\frac{1}{2}\,\beta^2_0 -
\frac{1}{2}\cdot\frac{2}{3}\,\Big(j_2 + \frac{1}{2}\,j_1 -
\frac{1}{2}\,\beta_0\Big)^2 +
\frac{1}{2}\cdot\frac{2}{3}\cdot\frac{1}{2^2}\,\beta^2_0\nonumber\\
\hspace{-0.5in}&& -\frac{1}{2}\cdot\frac{3}{4}\,\Big(j_3 +
\frac{2}{3}\, j_2 + \frac{2}{3}\cdot\frac{1}{2}\,j_1 -
\frac{1}{3}\,\beta_0\Big)^2 +
\frac{1}{2}\cdot\frac{3}{4}\cdot\frac{1}{3^2}\,\beta^2_0 -
\frac{1}{2}\cdot\frac{4}{5}\,\Big(j_4 + \frac{3}{4}\,j_3 +
\frac{3}{4}\cdot\frac{2}{3}\, j_2\nonumber\\
\hspace{-0.5in}&& + \frac{3}{4}\cdot\frac{2}{3}\cdot\frac{1}{2}\,j_1 -
\frac{1}{4}\,\beta_0\Big)^2 +
\frac{1}{2}\cdot\frac{4}{5}\cdot\frac{1}{4^2}\,\beta^2_0 -
\frac{1}{2}\cdot\frac{5}{6}\,\Big(j_5 + \frac{4}{5}\,j_4
+\frac{4}{5}\cdot \frac{3}{4}\,j_3 + \frac{4}{5}\cdot
\frac{3}{4}\cdot\frac{2}{3}\, j_2\nonumber\\
\hspace{-0.5in}&& +
\frac{4}{5}\cdot\frac{3}{4}\cdot\frac{2}{3}\cdot\frac{1}{2}\,j_1 -
\frac{1}{5}\,\beta_0\Big)^2 +
\frac{1}{2}\cdot\frac{5}{6}\cdot\frac{1}{5^2}\,\beta^2_0 -
\ldots - \frac{1}{2}\cdot\frac{k}{k+1}\,\Big(j_k +
\frac{k-1}{k}\,j_{k-1} \nonumber\\
\hspace{-0.5in}&& + \frac{k-1}{k}\cdot\frac{k-2}{k-1}\,j_{k-2} +
\ldots + \frac{k-1}{k}\cdot\frac{k-2}{k-1}\cdots \frac{2}{3}\,j_2 + \frac{k-1}{k}\cdot\frac{k-2}{k-1}\cdots
\frac{2}{3}\cdot\frac{1}{2}\,j_1 - \frac{1}{k}\,\beta_0\Big)^2 \nonumber\\
\hspace{-0.5in}&& +
\frac{1}{2}\cdot\frac{k}{k+1}\cdot\frac{1}{k^2}\,\beta^2_0\Bigg]\Bigg).
\end{eqnarray}
By performing $N$ integrations we obtain
\begin{eqnarray}\label{label6.26}
\hspace{-0.3in}&&\int\limits^{\infty}_{-\infty}d\beta_N
\int\limits^{\infty}_{-\infty}d\beta_{N-1}\ldots
\int\limits^{\infty}_{-\infty}d\beta_2
\int\limits^{\infty}_{-\infty}d\beta_1\,
\exp\Bigg(i\,\frac{I_{\perp}}{2\delta}\,
[(\beta_{N+1} - \beta_{N})^2 + (\beta_{N} - \beta_{N-1})^2\nonumber\\
\hspace{-0.3in}&& + \ldots +(\beta_2 - \beta_1)^2 + (\beta_1 -
\beta_0)^2 + 2\,j_N\,\beta_N + 2\,j_{N-1}\,\beta_{N-1} + \ldots + 2\,j_2\,
\beta_2 + 2\,j_1\,\beta_1 ]\Bigg) =\nonumber\\
\hspace{-0.3in}&&=\Bigg(\sqrt{\frac{2\pi
i\delta}{I_{\perp}}}\,\Bigg)^{N+1}\,\sqrt{\frac{I_{\perp}}{2\pi
i\Delta t}}\,
\exp\Bigg(i\,\frac{I_{\perp}}{2\Delta t}\,(
\beta_{N+1} - \beta_0)^2\Bigg)\nonumber\\
\hspace{-0.3in}&& 
\times\,\exp\Bigg(i\,\frac{I_{\perp}}{\delta}\,\beta_{N+1}\,
\Bigg(\frac{N}{N+1}\,j_N +
\frac{N}{N+1}\cdot\frac{N-1}{N}\,j_{N-1}
+\frac{N}{N+1}\cdot\frac{N-1}{N}\cdot\frac{N-2}{N-1}\,j_{N-2}\nonumber\\
\hspace{-0.5in}&& + \ldots +
\frac{N}{N+1}\cdot\frac{N-1}{N}\cdot\frac{N-2}{N-1}\cdots
\frac{2}{3}\,j_2 +
\frac{N}{N+1}\cdot\frac{N-1}{N}\cdot\frac{N-2}{N-1}\cdots
\frac{2}{3}\cdot\frac{1}{2}\,j_1\Bigg)\Bigg)\nonumber\\
\hspace{-0.3in}&&\times\,\exp\Bigg(i\,\frac{I_{\perp}}{\delta}\,
\Bigg[-\frac{1}{2}\cdot\frac{1}{2}\,\Big(j_1 - \beta_0\Big)^2 +
\frac{1}{2}\cdot\frac{1}{2}\,\beta^2_0 -
\frac{1}{2}\cdot\frac{2}{3}\,\Big(j_2 + \frac{1}{2}\,j_1 -
\frac{1}{2}\,\beta_0\Big)^2 +
\frac{1}{2}\cdot\frac{2}{3}\cdot\frac{1}{2^2}\,\beta^2_0\nonumber\\
\hspace{-0.3in}&& -\frac{1}{2}\cdot\frac{3}{4}\,\Big(j_3 +
\frac{2}{3}\, j_2 + \frac{2}{3}\cdot\frac{1}{2}\,j_1 -
\frac{1}{3}\,\beta_0\Big)^2 +
\frac{1}{2}\cdot\frac{3}{4}\cdot\frac{1}{3^2}\,\beta^2_0 -
\frac{1}{2}\cdot\frac{4}{5}\,\Big(j_4 + \frac{3}{4}\,j_3 +
\frac{3}{4}\cdot\frac{2}{3}\, j_2\nonumber\\
\hspace{-0.3in}&& + \frac{3}{4}\cdot\frac{2}{3}\cdot\frac{1}{2}\,j_1 -
\frac{1}{4}\,\beta_0\Big)^2 +
\frac{1}{2}\cdot\frac{4}{5}\cdot\frac{1}{4^2}\,\beta^2_0 -
\frac{1}{2}\cdot\frac{5}{6}\,\Big(j_5 + \frac{4}{5}\,j_4 +
\frac{4}{5}\cdot \frac{3}{4}\,j_3 + \frac{4}{5}\cdot
\frac{3}{4}\cdot\frac{2}{3}\, j_2\nonumber\\
\hspace{-0.3in}&& +
\frac{4}{5}\cdot\frac{3}{4}\cdot\frac{2}{3}\cdot\frac{1}{2}\,j_1 -
\frac{1}{5}\,\beta_0\Big)^2 +
\frac{1}{2}\cdot\frac{5}{6}\cdot\frac{1}{5^2}\,\beta^2_0 -
\ldots - \frac{1}{2}\cdot\frac{N}{N+1}\,\Big(j_N +
\frac{N-1}{N}\,j_{N-1} \nonumber\\
\hspace{-0.3in}&& + \frac{N-1}{N}\cdot\frac{N-2}{N-1}\,j_{N-2} +
\ldots + \frac{N-1}{N}\cdot\frac{N-2}{N-1}\cdots \frac{2}{3}\,j_2 +
\frac{N-1}{N}\cdot\frac{N-2}{N-1}\cdots
\frac{2}{3}\cdot\frac{1}{2}\,j_1
\nonumber\\
\hspace{-0.3in}&& - \frac{1}{N}\,\beta_0\Big)^2  +
\frac{1}{2}\cdot\frac{N}{N+1}\cdot\frac{1}{N^2}\,\beta^2_0\Bigg]\Bigg),
\end{eqnarray}
where we have replaced $(N+1)\,\delta = t_2 - t_1 = \Delta t$.

Now we can evaluate the derivatives with respect to $j_1, j_2, \ldots
,j_{N-1},j_N$. Due to the constraint $I_{\perp}/\delta \gg 1$ we can
keep only the leading order contributions in powers of
$I_{\perp}/\delta \gg 1$. The result reads
\begin{eqnarray}\label{label6.27}
\hspace{-0.5in}&&\frac{\partial}{\partial j_1}\frac{\partial}{\partial
j_2}\ldots \frac{\partial}{\partial j_{N-1}}\frac{\partial}{\partial
j_N} \int\limits^{\infty}_{-\infty}d\beta_N
\int\limits^{\infty}_{-\infty}d\beta_{N-1}\ldots
\int\limits^{\infty}_{-\infty}d\beta_2
\int\limits^{\infty}_{-\infty}d\beta_1\nonumber\\
\hspace{-0.5in}&&\times\,\exp\Bigg(i\,\frac{I_{\perp}}{2\delta}\,
[(\beta_{N+1} - \beta_{N})^2 + (\beta_{N} - \beta_{N-1})^2 + \ldots
+(\beta_2 - \beta_1)^2 + (\beta_1 - \beta_0)^2\nonumber\\
\hspace{-0.5in}&& + 2\,j_N\,\beta_N + 2\,j_{N-1}\,\beta_{N-1} + \ldots +
2\,j_2\, \beta_2 + 2\,j_1\,\beta_1 ]\Bigg)\Bigg|_{j_N = j_{N-1} = \ldots
=j_2 = j_1 = 0} =\nonumber\\
\hspace{-0.5in}&&= \Bigg(\sqrt{\frac{2\pi
i\delta}{I_{\perp}}}\,\Bigg)^{N+1}\,\sqrt{\frac{I_{\perp}}{2\pi
i\Delta t}}\, \exp\Bigg(i\,\frac{I_{\perp}}{2\Delta t}\,( \beta_{N+1}
- \beta_0)^2\Bigg)\,\Bigg(\frac{iI_{\perp}}{\delta}\Bigg)^N\,\nonumber\\
\hspace{-0.5in}&&\Bigg(\beta_{N+1}\,\frac{1}{N+1}+\beta_0\,\Bigg[\frac{1}{1\cdot 2} + \frac{1}{2\cdot 3} +
\frac{1}{3\cdot 4} + \ldots +  \frac{1}{N(N+1)}\Bigg]\Bigg)\,\nonumber\\
\hspace{-0.5in}&&\times\,\Bigg(\beta_{N+1}\,\frac{2}{N+1} + \beta_0\,\Bigg[\frac{2}{2\cdot 3} +
\frac{2}{3\cdot 4} + \ldots +  \frac{2}{N(N+1)}\Bigg]\Bigg)\,\nonumber\\
\hspace{-0.5in}&&\Bigg(\beta_{N+1}\,\frac{3}{N+1} +
\beta_0\,\Bigg[\frac{3}{3\cdot 4} + \ldots +
\frac{3}{N(N+1)}\Bigg]\Bigg)\,\nonumber\\
\hspace{-0.5in}&&\times\,\Bigg(\beta_{N+1}\,\frac{4}{N+1} +
\beta_0\, \Bigg[\frac{4}{4\cdot5 } +\ldots +
\frac{4}{N(N+1)}\Bigg]\Bigg)\ldots \Bigg(\beta_{N+1}\,\frac{N}{N+1} +
\beta_0\,\frac{N}{N(N+1)}\Bigg)=\nonumber\\
\hspace{-0.5in}&&= \Bigg(\sqrt{\frac{2\pi
i\delta}{I_{\perp}}}\,\Bigg)^{N+1}\,\sqrt{\frac{I_{\perp}}{2\pi
i\Delta t}}\, \exp\Bigg(i\,\frac{I_{\perp}}{2\Delta t}\,( \beta_{N+1}
- \beta_0)^2\Bigg)\,\Bigg(\frac{iI_{\perp}}{\delta}\Bigg)^N\,\nonumber\\
\hspace{-0.5in}&&\times\,\Bigg(\beta_{N+1}\,\frac{1}{N+1} +
\beta_0\,\frac{N}{N+1}\Bigg)\,\Bigg(\beta_{N+1}\,\frac{2}{N+1} +
\beta_0\,\frac{N-1}{N+1}\Bigg)\nonumber\\
\hspace{-0.5in}&&\times\,\Bigg(\beta_{N+1}\,\frac{3}{N+1} +
\beta_0\,\frac{N-2}{N+1}\Bigg)\,\times\,\ldots\times\,
\Bigg(\beta_{N+1}\,\frac{N}{N+1}
+ \beta_0\,\frac{1}{N+1}\Bigg)=\nonumber\\
\hspace{-0.5in}&&= \Bigg(\sqrt{\frac{2\pi
i\delta}{I_{\perp}}}\,\Bigg)^{N+1}\,\sqrt{\frac{I_{\perp}}{2\pi
i\Delta t}}\, \exp\Bigg(i\,\frac{I_{\perp}}{2\Delta t}\,( \beta_{N+1}
-  \beta_0)^2\Bigg)\,\Bigg(\frac{iI_{\perp}}{\delta}\Bigg)^N\,\nonumber\\
\hspace{-0.5in}&&\times\,\prod^{N}_{k=1}\Bigg(\beta_{N+1}\,\frac{k}{N+1} +
\beta_0\,\frac{N+1-k}{N+1}\Bigg).
\end{eqnarray}
Substituting Eq.(\ref{label6.27}) in Eq.(\ref{label6.24}) we obtain
the functional $F[I_{\perp},\beta_{N+1},\beta_0]$: 
\begin{eqnarray}\label{label6.28}
\hspace{-0.5in}&&F[I_{\perp},\beta_{N+1},\beta_0]
= \sqrt{\frac{I_{\perp}}{2\pi i\Delta t}}\,
\exp\Bigg(i\,\frac{I_{\perp}}{2\Delta t}\,( \beta_{N+1} -
\beta_0)^2\Bigg)\nonumber\\
\hspace{-0.5in}&&\times\,\lim_{\begin{array}{c} N\to \infty\\ \delta \to
0\end{array}}\Bigg(\frac{1}{8\pi}\Bigg)^N \Bigg(\sqrt{\frac{I_{\perp}}{2\pi i
\delta}}\,\Bigg)^{N+1}\,\prod^{N}_{k=1}\Bigg(\beta_{N+1}\,\frac{k}{N+1} +
\beta_0\,\frac{N+1-k}{N+1}\Bigg).
\end{eqnarray}
It is instructive to emphasize that this result can be obtained more easily if we sum up like terms in the exponent of the r.h.s. of
Eq.(\ref{label6.26}). This yields
\begin{eqnarray}\label{label6.29}
\hspace{-0.5in}&&\int\limits^{\infty}_{-\infty}d\beta_N
\int\limits^{\infty}_{-\infty}d\beta_{N-1}\ldots
\int\limits^{\infty}_{-\infty}d\beta_2
\int\limits^{\infty}_{-\infty}d\beta_1\,
\exp\Bigg(i\,\frac{I_{\perp}}{2\delta}\,
[(\beta_{N+1} - \beta_{N})^2 + (\beta_{N} - \beta_{N-1})^2\nonumber\\
\hspace{-0.5in}&& + \ldots +(\beta_2 - \beta_1)^2 + (\beta_1 -
\beta_0)^2 + 2\,j_N\,\beta_N + 2\,j_{N-1}\,\beta_{N-1} + \ldots + 2\,j_2\,
\beta_2 + 2\,j_1\,\beta_1 ]\Bigg) =\nonumber\\
\hspace{-0.5in}&&=\Bigg(\sqrt{\frac{2\pi
i\delta}{I_{\perp}}}\,\Bigg)^{N+1}\,\sqrt{\frac{I_{\perp}}{2\pi
i\Delta t}}\,
\exp\Bigg(i\,\frac{I_{\perp}}{2\Delta t}\,(
\beta_{N+1} - \beta_0)^2\Bigg)\nonumber\\
\hspace{-0.5in}&& \times\,\exp\Bigg(i\frac{I_{\perp}}{\delta}\Bigg[
\sum^{N}_{k=1}\Bigg(\beta_{N+1}\frac{k}{N+1} +
\beta_0\frac{N+1-k}{N+1}\Bigg)\,j_k -
\frac{1}{2}\sum^{N}_{n=1}\sum^{N}_{k=1}n (N+1 - k) j_n j_k\Bigg]\Bigg).
\end{eqnarray}
In order to understand the behaviour of the functional
$F[I_{\perp},\beta_{N+1},\beta_0]$ in the limit $N \to \infty$ we
suggest to evaluate the product
\begin{eqnarray}\label{label6.30}
{\Pi}[\beta_{N+1},\beta_0] = \prod^{N}_{k=1}\Bigg(\beta_{N+1}\,\frac{k}{N+1} +
\beta_0\,\frac{N+1-k}{N+1}\Bigg)
\end{eqnarray}
at $N \gg 1$ by using the $\zeta$--regularization. In the $\zeta$--regularization the evaluation of ${\Pi}[\beta_{N+1},\beta_0]$ runs the following way
\begin{eqnarray}\label{label6.31}
&&{\ell n}\,{\Pi}[\beta_{N+1},\beta_0] = \sum^{N}_{k=1}{\ell
n}\,\Bigg(\beta_{N+1}\,\frac{k}{N+1} +
\beta_0\,\frac{N+1-k}{N+1}\Bigg) = \nonumber\\
&&=\sum^{N}_{k=1}(-1)\,\frac{d}{ds}\Bigg(\beta_{N+1}\,\frac{k}{N+1} +
\beta_0\,\frac{N+1-k}{N+1}\Bigg)^{-s}\Bigg|_{s=0}=\nonumber\\
&&=-\frac{d}{ds}\sum^{N}_{k=1}\int\limits^{\infty}_0\frac{dz}{\Gamma(s)}\,
\exp\Bigg[-\Bigg(\beta_{N+1}\,\frac{k}{N+1} +
\beta_0\,\frac{N+1-k}{N+1}\Bigg)\,z\Bigg]\,z^{s-1}\Bigg|_{s=0}=\nonumber\\
&&=-\frac{d}{ds} \int\limits^{\infty}_0\frac{dz}{\Gamma(s)}\,
\left[\frac{\displaystyle e^{\textstyle -\beta_0\,z} - e^{\textstyle -
\beta_{N+1}\,z}}{\displaystyle 1 - \exp \Big( -\frac{\beta_{N+1} -
\beta_0}{N+1}\,z\Big)}- e^{\textstyle
-\beta_0\,z}\right]\,z^{s-1}\Bigg|_{s=0} =\nonumber\\ &&=-
(N+1)\,\frac{d}{ds} \int\limits^{\infty}_0\frac{dz}{\Gamma(s)}\,
\frac{\displaystyle e^{\textstyle -\beta_0\,z} - e^{\textstyle -
\beta_{N+1}\,z}}{\beta_{N+1} - \beta_0}\,z^{s-2}\Bigg|_{s=0}
=\nonumber\\ 
&&=- (N+1)\,\frac{d}{ds}\,\left[ \frac{1}{s-1}\,
\frac{\displaystyle \beta^{1-s}_0 - \beta^{1-s}_{N+1}}{\beta_{N+1} -
\beta_0}\right]_{s=0} =\nonumber\\
&&= - (N+1)\,\Bigg[1 - \frac{\beta_{N+1}\,{\ell
n}\,\beta_{N+1} - \beta_0\,{\ell n}\,\beta_0}{\beta_{N+1} - \beta_0}\Bigg]
=\nonumber\\
 &&=- (N+1)\,\frac{\displaystyle \beta_{N+1}\,{\ell
n}\,\frac{e}{\beta_{N+1}} - \beta_0\,{\ell
n}\,\frac{e}{\beta_0}}{\beta_{N+1} - \beta_0}.
\end{eqnarray}
Thus, the function ${\Pi}[\beta_{N+1},\beta_0]$ is defined by 
\begin{eqnarray}\label{label6.32}
{\Pi}[\beta_{N+1},\beta_0] = \exp\left( - (N+1)\,\frac{\displaystyle
\beta_{N+1}\,{\ell n}\,\frac{e}{\beta_{N+1}} - \beta_0\,{\ell
n}\,\frac{e}{\beta_0}}{\beta_{N+1} - \beta_0}\,\right),
\end{eqnarray}
where $e = 2.71828\ldots$. Due to the constraint $I_{\perp}/\delta
\gg 1$ the Euler angles $\beta_{N+1}$ and $\beta_0$ are less than
unity and the ratio in Eq.(\ref{label6.32}) is always positive
\begin{eqnarray}\label{label6.33}
\frac{\displaystyle
\beta_{N+1}\,{\ell n}\,\frac{e}{\beta_{N+1}} - \beta_0\,{\ell
n}\,\frac{e}{\beta_0}}{\beta_{N+1} - \beta_0} > 0.
\end{eqnarray}
The functional $F[I_{\perp},\beta_{N+1},\beta_0]$ is then defined by 
\begin{eqnarray}\label{label6.34}
\hspace{-0.5in}&&F[I_{\perp},\beta_{N+1},\beta_0]
= \sqrt{\frac{I_{\perp}}{2\pi i\Delta t}}\,
\exp\Bigg(i\,\frac{I_{\perp}}{2\Delta t}\,( \beta_{N+1} -
\beta_0)^2\Bigg)\nonumber\\
\hspace{-0.5in}&&\times\,\lim_{\begin{array}{c} N\to \infty\\ \delta \to
0\end{array}}\Bigg(\frac{1}{8\pi}\Bigg)^N \Bigg(\sqrt{\frac{I_{\perp}}{2\pi i
\delta}}\,\Bigg)^{N+1}\,\exp\left( - (N+1)\,\frac{\displaystyle
\beta_{N+1}\,{\ell n}\,\frac{e}{\beta_{N+1}} - \beta_0\,{\ell
n}\,\frac{e}{\beta_0}}{\beta_{N+1} - \beta_0}\,\right).
\end{eqnarray}
Thus $F[I_{\perp},\beta_{N+1},\beta_0]$ vanishes in the limit $N \to
\infty$. This result retains itself even if we change the
normalization factor of the evolution operator
\begin{eqnarray}\label{label6.35}
{\cal N} = \Bigg(\frac{I_{\perp}}{2\pi i
\delta}\sqrt{\frac{I_{\parallel}}{2\pi i \delta}}\,\Bigg)^{N+1} \to
(8\pi)^N\,\Bigg(\sqrt{\frac{I_{\parallel}}{I_{\perp}}}\,\Bigg)^{N+1} .
\end{eqnarray}
The renormalized functional $F[I_{\perp},\beta_{N+1},\beta_0]$,
defined by
\begin{eqnarray}\label{label6.36}
F[I_{\perp},\beta_{N+1},\beta_0]
&=& \sqrt{\frac{I_{\perp}}{2\pi i\Delta t}}\,
\exp\Bigg(i\,\frac{I_{\perp}}{2\Delta t}\,( \beta_{N+1} -
\beta_0)^2\Bigg) \nonumber\\
&\times&\lim_{N\to \infty}\exp\left( - (N+1)\,\frac{\displaystyle
\beta_{N+1}\,{\ell n}\,\frac{e}{\beta_{N+1}} - \beta_0\,{\ell
n}\,\frac{e}{\beta_0}}{\beta_{N+1} - \beta_0}\,\right),
\end{eqnarray}
vanishes in the limit $N\to \infty$ since the Euler angles
$\beta_{N+1}$ and $\beta_0$ are smaller compared with unity due to the
constraint $I_{\perp}/\delta \gg 1$ [3]. The vanishing of the
functional $F[I_{\perp},\beta_{N+1},\beta_0]$ in the limit $N \to
\infty$ agrees with our results obtained in Sects.\,4 and 5.

Substituting Eq.(\ref{label6.36}) in Eq.(\ref{label6.21}) we obtain
\begin{eqnarray}\label{label6.37}
Z_{\rm Reg}(R_{N+1},R_0) = 0.
\end{eqnarray}
This leads to the vanishing of the evolution operator $Z(R_2,R_1)$
given by Eq.(\ref{label6.1}) or Eq.(8) of Ref.[3], $Z(R_2,R_1) = 0$.

Thus, the evolution operator $Z(R_2,R_1)$ suggested 
in Ref.[3] for the description of Wilson loops in terms of
path integrals over gauge degrees of freedom is equal to zero
identically.  This agrees with our results obtained in Sects.\,4 and
5. As we have shown the vanishing of $Z(R_2,R_1)$ does not depend on
the specific regularization and discretization of the path
integral. In fact, this is an intrinsic property of the path integral
given by Eq.(\ref{label6.1}) that becomes obvious if the evaluation is
carried out correctly.

\section{Evolution
 operator $Z(R_2,R_1)$ and shift of energy levels of 
an axial--symmetric top} 
\setcounter{equation}{0}

\hspace{0.2in} In this Section we criticize the analysis of the evolution operator
$Z(R_2,R_1)$ carried out by Diakonov and Petrov via {\it the canonical
quantization of the axial--symmetric top} (see Eq.(12) of
Ref.[3]). Below we use the notations of Ref.[3].

The parallel transport operator 
\begin{eqnarray}\label{label7.1}
W_{\alpha\beta}(t_2,t_1) =
\Bigg[P\exp\Bigg(i\int\limits^{x(t_2)}_{x(t_1)}
A^a_{\mu}(x)\,T^a\,dx_{\mu}\Bigg)\Bigg]_{\alpha\beta}
=\Bigg[P\exp\Bigg(i\int\limits^{t_2}_{t_1}
A(t)\,dt\Bigg)\Bigg]_{\alpha\beta},
\end{eqnarray}
where $A(t) = A^a_{\mu}(x)\,T^a\,dx_{\mu}/dt$ is a tangent component of
the Yang--Mills field and $T^a\,(a=1,2,3)$ are the generators of
$SU(2)$ group in the representation $T$, has been reduced to the form
\begin{eqnarray}\label{label7.2}
W_{\alpha\beta}(t_2,t_1) = D^T_{\alpha\beta}(U(t_2)U^{\dagger}(t_1)),
\end{eqnarray}
(see Eq.(5) of Ref.[3]) due to the statement [3]: {\it The potential
$A(t)$ along a given curve can be always written as a ``pure gauge''}
\begin{eqnarray}\label{label7.3}
A_{\alpha\beta}(t) =i\,
D^T_{\alpha\gamma}(U(t))\,\frac{d}{dt}D^T_{\gamma\beta}(U^{\dagger}(t))
\end{eqnarray}
(see Eq.(4) of Ref.[3]).

By using  the parallel transport operator Eq.(\ref{label7.2}) the
Wilson loop $W_T(C)$ in the representation $T$ has been defined by
\begin{eqnarray}\label{label7.4}
W_T(C) = \sum_{\alpha}W_{\alpha\alpha}(t_2,t_1) =\sum_{\alpha}
D^T_{\alpha\alpha}(U(t_2)U^{\dagger}(t_1)).
\end{eqnarray}
(see Eq.(25) of Ref.[3]).  In terms of the evolution operator
$Z(R_2,R_1)$ given by Eq.(\ref{label6.1}) (see Eq.(8) of Ref.[3]) the
parallel transport operator $W_{\alpha\beta}(t_2,t_1)$ has been
recast into the form
\begin{eqnarray}\label{label7.5}
W^{\rm DP}_{\alpha\beta}(t_2,t_1) =\int\!\!\!\int dR_1\,dR_2\sum_{T', m}(2T' +
1)\, D^{T'}_{\alpha
m}(U(t_2)R^{\dagger}_2)D^{T'}_{m\beta}(R_1U^{\dagger}(t_1))\,Z(R_2,R_1),
\end{eqnarray}
where the index ${\rm DP}$ means that the parallel transport operator
is taken in the Diakonov--Petrov (DP) representation. The Wilson loop
$W^{\rm DP}_T(C)$ in the DP--representation reads
\begin{eqnarray}\label{label7.6}
W^{\rm DP}_T(C) = \int\!\!\!\int dR_1\,dR_2\sum_{T', m, \alpha}(2T' +
1)\, D^{T'}_{\alpha
m}(U(t_2)R^{\dagger}_2)D^{T'}_{m\alpha}(R_1U^{\dagger}(t_1))\,Z(R_2,R_1).
\end{eqnarray}
Of course, if the DP--representation were correct we should get $W^{\rm
DP}_T(C) = W_T(C)$, where $ W_T(C)$ is determined by
Eq.(\ref{label7.4}).

 The regularized evolution operator $Z_{\rm Reg}(R_2,R_1)$ given by
 Eq.(\ref{label6.3}) (see also Eq.(9) of Ref.[3]) can be represented
 in the form of {\it a sum over possible intermediate states},
 i.e. eigenfunctions of the axial--symmetric top
\begin{eqnarray}\label{label7.7}
Z_{\rm Reg}(R_2,R_1) =\sum_{J, m, k}(2J + 1)\, D^J_{m
k}(R_2)D^J_{k m}(R^{\dagger}_1)\,e^{\textstyle -i(t_2-t_1)\,E_{J m}},
\end{eqnarray}
(see Eq.(12) of Ref.[3]), where $E_{J m}$ are the eigenvalues of the
Hamiltonian of the axial--symmetric top 
\begin{eqnarray}\label{label7.8}
E_{J m} = \frac{J(J+1) - m^2}{2 I_{\perp}} + \frac{(m - T)^2}{2
I_{\parallel}}
\end{eqnarray}
(see Eq.(11) of Ref.[3]).

As has been stated in Ref.[3]: {\it If we now take to zero
$I_{{\perp},{\parallel}} \to 0$ (first $I_{\parallel}$, then
$I_{\perp}$) we see that in the sum (12) only the lowest energy
intermediate state survives with $m=J=T$. The resulting phase factor
from the lowest energy state can be absorbed in the normalization
factor in eq.(9) since that corresponds to a shift in the energy
scale.}

The statement concerning the possibility to absorb the fluctuating
factor $\exp[-i(t_2-t_1)\,T/2 I_{\perp}]$ in the normalization 
of the path integral representing the evolution operator is the main
one allowing the r.h.s. of Eqs.(\ref{label7.5}) and (\ref{label7.6})
to escape from the vanishing in the limit $I_{\perp}\to 0$. 

In reality such a removal of the fluctuating factor is prohibited
since this leads to the change of the starting symmetry of
the system from $SU(2)$ to $U(2)$. In order to make this more 
transparent we suggest to insert $Z_{\rm Reg}(R_2,R_1)$ 
of Eq.(\ref{label7.7}) into Eq.(\ref{label7.6}) and to
express the Wilson loop $W^{\rm DP}_T(C)$ in terms of {\it a sum
over possible intermediate states}, the eigenfunctions of the
axial--symmetric top. The main idea of this substitution is the
following: as the Wilson loop is a physical quantity which can be
measured, all irrelevant normalization factors should be canceled for
the evaluation of it. Therefore, if the oscillating factor
$\exp[-\,i\,(t_2 - t_1)\,T/2I_{\perp}]$ can be really removed by a
renormalization of something, the Wilson loop should not depend on
this factor.

Substituting Eq.(\ref{label7.7}) in Eq.(\ref{label7.6}) and
integrating over $R_1$ and $R_2$ we obtain the following expansion
for the parallel transport operator in the DP--representation
\begin{eqnarray}\label{label7.9}
W^{\rm DP}_{\alpha\beta}(t_2,t_1)
=\sum_{T'}\sum^{T'}_{m=-T'}
D^{T'}_{\alpha\beta}(U(t_2)U^{\dagger}(t_1))\,e^{\textstyle
-i(t_2-t_1)\,E_{T' m}}.
\end{eqnarray}
Setting $\alpha = \beta$ and summing over $\alpha$ we get the
DP--representation for Wilson loops
\begin{eqnarray}\label{label7.10}
W^{\rm DP}_T(C) = \sum_{\alpha}W^{\rm DP}_{\alpha\alpha}(t_2,t_1)
=\sum_{T'}\sum^{T'}_{m=-T'} D^{T'}_{\alpha\alpha}(U(t_2) 
U^{\dagger}(t_1))\,e^{\textstyle - i(t_2-t_1)\,E_{T'  m}}.
\end{eqnarray}
Due  to the definition (\ref{label7.4}) the r.h.s. of
Eq.(\ref{label7.10}) can be rewritten in the form 
\begin{eqnarray}\label{label7.11}
W^{\rm DP}_T(C) = \sum_{\alpha}W^{\rm DP}_{\alpha\alpha}(t_2,t_1)
=\sum_{T'}\sum^{T'}_{m=-T'}W_{T'}(C)\,e^{\textstyle -i(t_2-t_1)\,E_{T'
m}}, 
\end{eqnarray}
where $W_{T'}(C)$ is the Wilson loop in the $T'$ representation
determined by Eq.(\ref{label7.4}).  The relation
(\ref{label7.11}) agrees to some extent with our expansion given
by Eq.(\ref{label7.1}).

Following Ref.[3] and taking the limit $I_{\parallel} \to 0$ we obtain $m
= T$. This reduces the r.h.s. of Eq.(\ref{label7.11}) to the form
\begin{eqnarray}\label{label7.12}
W^{\rm DP}_T(C) = \sum_{T'}W_{T'}(C)\,\exp\Bigg[ -i(t_2-t_1)\,\frac{T'(T' + 1)
- T^2}{2 I_{\perp}}\Bigg].
\end{eqnarray}
Now according to the prescription of Ref.[3] we should take the limit
$I_{\perp} \to 0$. Following again Ref.[3] and setting $T' = T$ we arrive at
the relation
\begin{eqnarray}\label{label7.13}
W^{\rm DP}_T(C) &=& W_T(C)\,\exp\Bigg[ -i(t_2-t_1)\,\frac{T}{2
I_{\perp}}\Bigg].
\end{eqnarray}
Since the average value of the Wilson loop is an observable quantity
and any averaging over gauge fields does not affect the oscillating
factor $\exp[ i(t_2-t_1)\,T/2
I_{\perp}]$, one can write
\begin{eqnarray}\label{label7.14}
\langle W^{\rm DP}_T(C)\rangle = \langle W_T(C)\rangle
\,\exp\Bigg[-i(t_2-t_1)\,\frac{T}{2I_{\perp}}\Bigg].
\end{eqnarray}
According to the Wilson's criterion of confinement Eq.(\ref{label1.8})
one should set
\begin{eqnarray}\label{label7.15}
\langle W_T(C)\rangle = e^{\textstyle - \sigma\,{\cal A}},
\end{eqnarray}
where $\sigma$ and ${\cal A}$ are a string tension and a minimal area,
respectively.

Substituting Eq.(\ref{label7.15}) in Eq.(\ref{label7.14}) one arrives
at the expression
\begin{eqnarray}\label{label7.16}
\langle W^{\rm DP}_T(C)\rangle = e^{\textstyle - \sigma\,{\cal A}}
\,\exp\Bigg[-i(t_2-t_1)\,\frac{T}{2I_{\perp}}\Bigg].
\end{eqnarray}
It seems to be rather obvious that the r.h.s. of Eq.(\ref{label7.16})
tends to zero due to a strongly oscillating factor, i.e. 
\begin{eqnarray}\label{label7.17}
\langle W^{\rm DP}_T(C)\rangle_{\rm Reg} &=& \lim_{I_{\perp}\to
0}\langle W^{\rm DP}_T(C)\rangle = \lim_{I_{\perp}\to 0} e^{\textstyle
- \sigma\,{\cal A}}
\,\exp\Bigg[-i(t_2-t_1)\,\frac{T}{2I_{\perp}}\Bigg] =\nonumber\\
&=&e^{\textstyle - \sigma\,{\cal A}} \lim_{I_{\perp}\to 0}
\,\exp\Bigg[-i(t_2-t_1)\,\frac{T}{2I_{\perp}}\Bigg] = 0.
\end{eqnarray}
Really, there is no quantity that can absorb this factor.

The only possibility to remove the oscillating factor $\exp[
-i(t_2-t_1)\,T/2 I_{\perp}]$ is to absorb this phase factors in the
matrices $U(t_2)$ and $U^{\dagger}(t_1)$ which describe the degrees of
freedom of the gauge potential $A(t)$ via relation
(\ref{label7.2}). In this case Eq.(\ref{label7.13}) can given by
\begin{eqnarray}\label{label7.18}
&&W^{\rm DP}_T(C) =  W_T(C)\,\exp\Bigg[ -i(t_2-t_1)\,\frac{T}{2
I_{\perp}}\Bigg]= \nonumber\\
&&=\sum_{\alpha}
D^T_{\alpha\alpha}(U(t_2)U^{\dagger}(t_1))\,\exp\Bigg[
-i(t_2-t_1)\,\frac{T}{2 I_{\perp}}\Bigg] = \sum_{\alpha}
D^T_{\alpha\alpha}(\bar{U}(t_2)\bar{U}^{\dagger}(t_1)),
\end{eqnarray}
where we have denoted 
\begin{eqnarray}\label{label7.19}
\bar{U}(t_2) &=& U(t_2)\,e^{\textstyle i\,t_2\,T/2
I_{\perp}},\nonumber\\ \bar{U}^{\dagger}(t_1) &=&
U^{\dagger}(t_1)\,e^{\textstyle -\,i\,t_1\,T/2 I_{\perp}}.
\end{eqnarray}
However, the matrices $\bar{U}(t_2)$ and $\bar{U}^{\dagger}(t_1)$ are
now elements of the group $U(2)$ instead of $SU(2)$. Thus, the
shift of the energy level of the ground state of the axial--symmetric
top suggested by Diakonov and Petrov in order to remove the
oscillating factor changes crucially the starting symmetry of the
theory from $SU(2)$ to $U(2)$.  Since the former is
not allowed the oscillating factor $\exp[ -i(t_2-t_1)\,T/2 I_{\perp}]$
cannot be removed. As a result in the limit $I_{\perp} \to 0$ we
obtain
\begin{eqnarray}\label{label7.20}
\langle W^{\rm DP}_T(C)\rangle_{\rm Reg} &=& \lim_{I_{\perp}\to
0}\langle W^{\rm DP}_T(C)\rangle = 0.
\end{eqnarray}
The vanishing of Wilson loops in the DP--representation agrees with
our results obtained in Sects.\,4, 5 and 6 and 
confirms our claim that this path integral representation of
Wilson loops is incorrect.

\section{The non--Abelian Stokes theorem}
\setcounter{equation}{0}

\hspace{0.2in} The derivation of the  area--law falloff promoted great interests
in  the non--Abelian Stokes theorem expressing the
exponent of Wilson loops in terms of a  surface integral over the
 2--dimensional surface $S$ with the  boundary $C = \partial S$ [20]
\begin{eqnarray}\label{label8.1}
\hspace{-0.2in}{\rm tr}\,{\cal P}_Ce^{\textstyle i\,g\,\oint_C d
x_{\mu}\,A_{\mu}(x)} = {\rm tr}\,{\cal P}_S\,
e^{\textstyle  i\,g\,\frac{1}{2}
\int\!\!\!\int\limits_S 
d\sigma_{\mu\nu}(y)\,U(C_{xy})\,G_{\mu\nu}(y)\,U(C_{yx})},
\end{eqnarray}
where ${\cal P}_S$ is the surface ordering operator [20],
$d\sigma_{\mu\nu}(y)$ is a 2--dimensional surface element in
4--dimensional space--time, $x$ is a current point on the contour $C$,
i.e. $x \in C$, $y$ is a point on the surface $S$, i.e. $y \in S$, and
$G_{\mu\nu}(y) = \partial_{\mu} A_{\nu}(y) - \partial_{\nu} A_{\mu}(y)
- ig[A_{\mu}(y), A_{\nu}(y)]$ is the field strength tensor. The
procedure for the derivation of the non--Abelian Stokes theorem in the
form of Eq.(\ref{label8.1}) contains a summation of contributions of
closed paths around infinitesimal areas and these paths are linked to
the reference point $x$ on  the contour $C$  via 
parallel transport operators.  The existence of closed paths linked to
the references point $x$ on the contour $C$ is a {\it necessary} and a
{\it sufficient} condition for the derivation of the non--Abelian
Stokes theorem Eq.(\ref{label8.1}).

Due to the absence of closed paths it is rather clear that the path
integral representation for Wilson loops cannot be applied to the
derivation of the non--Abelian Stokes theorem.  In fact, the
evaluation of the path integral over gauge degrees of freedom demands
the decomposition of the closed contour $C$ into a set of
infinitesimal segments which can be never closed.  Let us prove this
statement by assuming the converse. Suppose that by representing the
path integral over gauge degrees of freedom in the form of the
$n$--dimensional integral (\ref{label2.7}) we have a closed
segment.  Let the segment $C_{x_k x_{k-1}}$ be closed and the point
$x'$ belong to the segment $C_{x_k x_{k-1}}$, $x' \in C_{x_k
x_{k-1}}$. By using Eq.(\ref{label2.2}) we can represent the character
$\chi[U^{\Omega}_r(C_{x_k x_{k-1}})]$ by
\begin{eqnarray}\label{label8.2}
\hspace{-0.2in}&&\chi[U^{\Omega}_r(C_{x_k x_{k-1}})] =
\chi[U^{\Omega}_r(C_{x_i x'})U^{\Omega}_r(C_{x' x_{i-1}})]
=\nonumber\\
\hspace{-0.2in}&&=\chi[\Omega(x_i)U_r(C_{x_i x'})U_r(C_{x'
x_{i-1}})\Omega(x_{i-1})] = \nonumber\\ 
\hspace{-0.2in}&&=d_r\int
D\Omega_r(x')\chi[\Omega_r(x_i)U_r(C_{x_i
x'})\Omega^{\dagger}_r(x')]\,\chi[\Omega_r(x') U_r(C_{x'
x_{i-1}})\Omega(x_{i-1})]=\nonumber\\ 
\hspace{-0.2in}&&=d_r\int
D\Omega_r(x')\chi[U^{\Omega}_r(C_{x_i
x'})]\,\chi[U^{\Omega}_r(C_{x' x_{i-1}})].
\end{eqnarray}
This transforms a $(n-1)$--dimensional integral with one closed
infinitesimal segment into a $n$--dimensional integral without closed
segments. Since finally $n$ tends to infinity there is no closed
segments for the representation for the path integral in the form of a
$(n-1)$--dimensional integral. As this statement is general and valid
for any path integral representation of Wilson loops, so one can
conclude that no further non--Abelian Stokes theorem can be derived
within any path integral approach to Wilson loops.

\section{Conclusion}
\setcounter{equation}{0}

\hspace{0.2in} By using well defined properties of group characters we have shown
that the path integral over gauge degrees of freedom of the Wilson
loop which was used in Eq.(2.13) of Ref.[11] for a lattice evaluation
of the average value of Wilson loops can be derived in continuum
space--time in non--Abelian gauge theories with the gauge group
$SU(N)$. The resultant integrand of the path integral contains a phase
factor which is not projected onto Abelian degrees of freedom of
non--Abelian gauge fields and differs substantially from the
representation given in Ref.[3]. The important point of our
representation is the summation over all states of the given
irreducible representation $r$ of $SU(N)$. For example, in $SU(2)$ the
phase factor is summed over all values of the colourmagnetic quantum
number $m_j$ of the irreducible representation $j$ of colour
charges. This contradicts Eq.(23) of Ref.[3], where only term with the
highest value of the colourmagnetic quantum number $m_j = j$ are taken
into account and the other $2j$ terms are lost. This loss is caused by
an artificial regularization procedure applied in Ref.[3] for the
definition of the path integral over gauge degrees of freedom.

As has been stated by Diakonov and Petrov in Ref.[3] the path integral
over gauge degrees of freedom representing Wilson loops {\it is not of
the Feynman type, therefore, it depends explicitly on how one
``understands'' it, i.e. how it is discretized and regularized}. In
order to {\it understand} the path integral over gauge degrees of
freedom Diakonov and Petrov [3] suggested a regularization procedure
drawing an analogy between gauge degrees of freedom and dynamical
variables of the axial--symmetric top with moments of inertia
$I_{\perp}$ and $I_{\parallel}$.  The final expression for the path
integral of the Wilson loop has been obtained in the limit $I_{\perp},
I_{\parallel} \to 0$.  

In order to make the incorrectness of this expression more transparent
we have evaluated the path integral for specific gauge field
configurations (i) a pure gauge field and (ii) $Z(2)$ center vortices
with spatial azimuthal symmetry.  The direct evaluation of path
integrals representing Wilson loops for these gauge field
configurations has given the value zero for both cases. These results do
not agree with the correct values.

One can show that Eq.(\ref{label5.9}) can be
generalized for any contour of a Wilson loop in $SU(2)$
\begin{eqnarray}\label{label9.1}
W_{1/2}(C)&=&\int \prod_{x\in C} D \Omega(x)\,e^{\textstyle ig
\oint_{C}dx_{\mu}\,{\rm tr}[t^3 A^{\Omega}_{\mu}(x)]} = \nonumber\\
&=&\lim_{n\to \infty}\sum_{j}\Bigg[\frac{a_j}{2j+1}\Bigg]^n(2j +
1)W_j(C) = 0,
\end{eqnarray}
where $W_j(C)$ in the r.h.s. is defined by Eq.(\ref{label2.1}) in
terms of the path--ordering operator ${\cal P}_C$. Further, the result
(\ref{label9.1}) can be extended to any irreducible representation
of $SU(2)$.

This statement we have supported by a direct evaluation of the
evolution operator $Z_{\rm Reg}(R_2,R_1)$ defined by Eq.(14) of
Ref.[3], representing the assumption by Diakonov and Petrov for Wilson
loops in terms of the path integral over gauge degrees of freedom. As
we have shown in Sect.\,6 the regularized evolution operator $Z_{\rm
Reg}(R_2,R_1)$, evaluated correctly, is equal to zero. This agrees
with our results obtained in Sects.\,4 and 5. In Sect.\,7 we have
shown that the removal of the oscillating factor from the evolution
operator suggested in Ref.[3] via a shift of energy levels of the
axial--symmetric top is prohibited. Such a shift of energy levels
leads to a change of the starting symmetry of the system from $SU(2)$
to $U(2)$. By virtue of the oscillating factor the Wilson loop
vanishes in the limit $I_{\parallel}, I_{\perp} \to 0$ in agreement
with our results in Sects.\,4, 5 and 6.

We hope that the considerations in Sects.\,4--7 are more than
 enough to persuade even the most distrustful reader that the path
 integral representation for the Wilson derived by Diakonov and Petrov
 by means of {\it special regularization and understanding} of the
 path integral over gauge degrees of freedom is erroneous.

The use of an erroneous path integral representation for Wilson
loops in Ref.[7] has led to the conclusion that for large distances the
average value of Wilson loops shows area--law
falloff for any irreducible representation $r$ of $SU(N)$. 
Unfortunately, this result is not supported by numerical
simulations of lattice QCD [8]. At large distances, colour charges
with non--zero $N$--ality have string tensions of the corresponding
fundamental representation, whereas colour charges with zero
$N$--ality are screened by gluons and cannot form a string. Therefore,
the result obtained in Ref.[7] cannot be considered as {\it
a new check of confinement in lattice calculations} as has been argued
by the authors of Ref.[7].

We would like to accentuate that the problem we have touched in this
paper is not of marginal interest and a path integral, if derived by
means of an unjustified regularization procedure, would hardly compute
the same physical number as the correct one. We argue that no
regularization procedure can lead to specific dynamical
constraints. In fact, the regularization procedure drawing the analogy
with the axial--symmetric top has led to the result supporting the
hypothesis of Maximal Abelian Projection pointed out by 't Hooft
[14]. Any proof of this to full extent dynamical hypothesis through a
regularization procedure and through specific {\it understanding} of
the path integral should have seemed dubious and suspicious.

Finally, we have shown that within any path integral representation for
Wilson loops in terms of gauge degrees of freedom no non--Abelian
Stokes theorem in addition to Eq.(\ref{label8.1}) can be derived.
Indeed, the Stokes theorem replaces a line integral over a closed
contour by a surface integral with the closed contour as the boundary
of a surface. However, approximating the path integral by an
$n$--dimensional integral at $n \to \infty$ there are no closed paths
linking two adjacent points along Wilson loops.  Thereby, the line
integrals over these open paths cannot be replaced by surface
integrals. Thus, we argue that any non-Abelian Stokes theorem can be
derived only within the definition of Wilson loops through the path
ordering procedure (\ref{label8.1}). Of course, one can represent
the surface--ordering operator ${\cal P}_S$ in Eq.(\ref{label8.1}) in
terms of a path integral over gauge degrees of freedom, but this
should not be a new non--Abelian Stokes theorem in comparison with the
old one given by Eq.(\ref{label8.1}). That is why the claims of
Ref.[3--5] concerning new versions of the non--Abelian Stokes theorems
derived within path integral representations for Wilson loops seem
unjustified.

\newpage

\section{Appendix. Comments on hep--lat/0008004 by Diakonov and Petrov}
\setcounter{equation}{0}
\subsection{None non--Abelian Stokes theorem can be derived within path 
integral representation of the Wilson loop over gauge degrees of freedom}
\setcounter{equation}{0}

\hspace{0.2in} In a set of publications [3,7,15,21,22] Diakonov and
Petrov have pointed the possibility to derive a non--Abelian Stokes
theorem for the Wilson loop represented by a path integral over gauge
degrees of freedom. This statement contradicts to a theorem that have
been proved recently by Faber {\it et al.} [23] (see also Sect.\,8 of
this manuscript).  The main point of this theorem has been proved by
using well--defined properties of group characters and refers to the
well--known result stating that a linear integral over a contour $C$
can be replaced by a surface integral (the Stokes theorem) only if the
contour $C$ is closed. As has been shown in [23] such a necessary
condition of the application of the Stokes theorem is violated for the
Wilson loop represented by a path integral over gauge degrees of
freedom. Below we repeat our statement.

In a non--Abelian gauge theory the Wilson loop defined for an
irreducible representation $r$ of the $SU(N)$ gauge group can be
represented in the form of a path integral over gauge degrees of
freedom as follows [23] (see Eq.(16) of Ref.[23])
\begin{eqnarray}\label{label10.1}
W_r(C) = \frac{1}{d^2_r}\int \prod_{x\in
C}[d_rD\Omega_r(x)]\,\chi_r[U^{\Omega}_r(C_{xx})],
\end{eqnarray}
where $d_r$ is a dimension of the irreducible representation $r$,
$\Omega_r(x)$ is the gauge function in the representation $r$ and
$\chi_r[U^{\Omega}_r(C_{xx})]$ is the group character defined for the
irreducible representation $r$, $C_{xx}$ is a Wilson contour. Then,
$U^{\Omega}_r(C_{xx})$ is given by
\begin{eqnarray}\label{label10.2}
U_r(C_{xx}) = {\cal P}_Ce^{\textstyle i\,g\,\oint_{C_{xx}} d
z_{\mu}\,A^{(r)}_{\mu}(z)},
\end{eqnarray}
where $A^{(r)}_{\mu}(x)$ is a gauge field in irreducible
representation $r$ of $SU(N)$ gauge group, ${\cal P}_C$ is the
operator ordering colour matrices along the path $C$. Then, the
quantity $U^{\Omega}_r(C_{xx})$ is defined by [5]
\begin{eqnarray}\label{label10.3}
U^{\Omega}_r(C_{xx}) = \Omega_r(x) U_r(C_{xx}) \Omega^{\dagger}_r(x),
\end{eqnarray}
For the matrices $U_r(C_{yx})$ defined for the open contour $C_{yx}$
linking two points $x$ and $y$ one has
\begin{eqnarray}\label{label10.4}
U_r(C_{yx}) = {\cal P}_{C_{yx}}e^{\textstyle i\,g\,\int_{C_{yx}} d
z_{\mu}\,A^{(r)}_{\mu}(z)}
\end{eqnarray}
and
\begin{eqnarray}\label{label10.5}
U^{\Omega}(C_{y x}) = \Omega(y)\,U(C_{y x})
\,\Omega^{\dagger}(x),
\end{eqnarray}
respectively.

The evaluation of the path integral over gauge functions
$\Omega_r(x)$ can be carried out only via
the discretization of the path integral (\ref{label10.1}). Such a
discretization can be unambiguously performed by using well--defined
properties of group characters. The discretized form reads [23] (see
Eq.(15) of Ref.[23])

\begin{eqnarray}\label{label10.6}
\hspace{-0.5in}&&W_r(C)= \frac{1}{d^2_r}\,\lim_{n\to \infty}\int
D\Omega_r(x_n)\,\chi_r[U^{\Omega}_r(C_{x_n x_{n-1}})]\int
D\Omega_r(x_{n-1})\,\chi_r[U^{\Omega}_r(C_{x_{n-1}
x_{n-2}})]\ldots\nonumber\\ \hspace{-0.5in}&&\times \int
D\Omega_r(x_2)\,\chi_r[U^{\Omega}_r(C_{x_2 x_1})]\int
D\Omega_r(x_1)\,\chi_r[U^{\Omega}_r(C_{x_1 x_n})] =\nonumber\\
\hspace{-0.5in}&&=\frac{1}{d^2_r}\,\lim_{n\to \infty}\int
D\Omega_r(x_n)\,\chi_r[\Omega_r(x_n)U_r(C_{x_n
x_{n-1}})\Omega^{\dagger}_r(x_{n-1})]\nonumber\\
\hspace{-0.5in}&&\times\int
D\Omega_r(x_{n-1})\,\chi_r[\Omega_r(x_{n-1})U^{\Omega}_r(C_{x_{n-1}
x_{n-2}})\Omega^{\dagger}_r(x_{n-2})]\ldots\nonumber\\
\hspace{-0.5in}&&\times\int
D\Omega_r(x_2)\,\chi_r[\Omega_r(x_2)U_r(C_{x_2
x_1})\Omega^{\dagger}_r(x_1)]\int
D\Omega_r(x_1)\,\chi_r[\Omega_r(x_1)U_r(C_{x_1x_n})\Omega^{\dagger}_r(x_n)],
\end{eqnarray}
where $C_{x_k x_{k-1}}$ are open infinitesimal segments linking two
adjoining points $x_{k-1}$ and $x_k$ and the relations
$U^{\Omega}_r(C_{x_k x_{k-1}}) = \Omega_r(x_k) U_r(C_{x_k x_{k-1}})
\Omega^{\dagger}_r(x_{k-1})$ have been used [23].

Since infinitesimal contours $C_{x_k x_{k-1}}$ are open and the integrations over $\Omega_r(x_k)$ are independent of the integration over $\Omega_r(x_{k-1})$
, the necessary
condition of the application of the Stokes theorem, i.e. a replacement
of a linear integral over closed contour by a surface integral over an
area embraced by the contour, is violated. This implies that none
non--Abelian Stokes can be derived within the framework of a path
integral representation of the Wilson in terms of gauge degrees of
freedom.

According to this theorem any non--Abelian Stokes theorem derived by
Diakonov and Petrov in Refs.[3,7,15,21,22] within path integral
representation of the Wilson loop over gauge degrees of freedom is
very much suspicious and is doomed to be erroneous.

We argue that in any non--Abelian gauge theory the non--Abelian Stokes
theorem within the standard definition of the
Wilson loop can only be derived via the path--ordering operator [20]. This non--Abelian Stokes
theorem is unique and no other exists.

\subsection{Path integral representation of the Wilson loop suggested 
by Diakonov and Petrov is erroneous}

\hspace{0.2in} In the recent manuscript [15] Diakonov and Petrov have
claimed that the path integrals over gauge degrees of freedom
representing in $SU(2)$ gauge theory Wilson loops defined for the pure
gauge field and the $Z(2)$ center vortex with spatial azimuthal
symmetry have been incorrectly calculated in our recent paper
[23]. The key point of this claim is the incompleteness of the
functions $\chi_j[t^3U]$, where $U$ is an element of $SU(2)$, $U\in
SU(2)$, and $t^3$ is the matrix representation of the third generator
of $SU(2)$ in the irreducible representation $j$.  Diakonov and Petrov
suggested another expansion obeying the completeness condition. As has
been shown in Sects.\,4 and 5 our results for the Wilson loop
represented by the path integral over gauge degrees of freedom
suggested by Diakonov and Petrov are retained for the correct
expansion. Thus, we conclude that in [15] Diakonov and Petrov have not
succeeded in refuting our statement that their representation for the
Wilson loop suggested in Ref.[3] is erroneous and cannot be used for a
new check of confinement [7].

\subsection{Evolution operator $Z_{\rm Reg}(R_2,R_1)$ is calculated by 
Diakonov and Petrov incorrectly} 

First, we would like to cite Diakonov and Petrov (p.11 of Ref.[15]):"In
section 6 of their paper FITZ\footnote{This is abbreviation from Faber,
Ivanov, Troitskaya, Zach used by Diakonov and Petrov} attempt to
compute the regularized evolution operator for the "Wess--Zumino"
action, following directly our approach. This calculation has been
presented in some detail in original paper [3], however, FITZ seem to
be dissatisfied by it and present their own. Their final answer
(eq.(98) and (112) of ref.[11]), which differs from our, is a result
of several mistakes.

First, going from eq.(98) to eq.(87) FITZ use a strange relation,
$$
\exp\Bigg(\sum^{N}_{n=0}(-i)\,\frac{I_{\perp}}{2\delta}(-4)\Bigg) =
\exp\Bigg(iN(N+1)\,\frac{I_{\perp}}{\delta}\Bigg),\eqno(37)
$$
instead of the correct (and trivial) $\exp(i2(N+1)I_{\perp}/\delta)$,
where $I_{\perp}$ and $\delta$ are constants and $N$ is the number of
pieces in which one divides the contour.

Second, and more important, both equations in (91) are erroneous, they
do not follow from eq.(89) from where they are derived. Passing from
eq.(89) to eq.(91) one gets:
$$
{\rm Tr}(R_nR^{\dagger}_{n+1}) = 2 - \frac{1}{4}[\delta\alpha^2_n +
\delta \beta^2_n + \delta\gamma^2_n +
2\delta\alpha_n\delta\gamma_n\,\cos\beta_n],\eqno(38)
$$
$$
{\rm Tr}(R_nR^{\dagger}_{n+1}\tau_3) = i(\delta\alpha_n +
\delta\gamma_n\,\cos\beta_n),
$$
$$
\delta\alpha_n = \alpha_{n+1} - \alpha_n, \quad \delta\beta_n =
\beta_{n+1} - \beta_n, \quad \delta\gamma_n = \gamma_{n+1} - \gamma_n.
$$
FITZ have written these formulae without the crucial factor $
\cos\beta_n$. Because of this mistake the subsequent integration over
Euler angles $ \alpha, \beta, \gamma$ becomes Gaussian, and the
evolution operator for the axial top, as computed by FITZ, in fact
becomes that a free particle, which is definitely wrong. The factors $
cos\beta_n$ being reinstalled, the derivation returns to that of our
paper [1]."

The factor Eq.(37) discussed by Diakonov and Petrov does not effect
the final result and cancels itself finally. It is rather clear that
this "mistake" is nothing more than a trivial misprint. Unfortunately,
there are some misprints in our paper [23]. For example in Eqs.(49) and
(74) one has to read $W_{1/2}(C)$ and $W_{1/2}(\rho)$ instead of
$W_{J}(C)$ and $W_{J}(\rho)$, respectively. In Appendix one should
read $0 \le \alpha \le 2\pi, 0\le \gamma \le 2\pi$ and $0 \le \beta
\le \pi$ instead of $0 \ge \alpha \ge 2\pi, 0\ge \gamma \ge 2\pi$ and
$0 \ge \beta \ge \pi$.

Before, the explanation of the "Second remark" we would like to
attract attention of readers to the manner of the expounding of the
problem accepted by Diakonov and Petrov.

They write "Second and more important, both equations (91) are
erroneous, they do not follow from eq.(89) from where they are
derived."

This sentence makes an oritation of the reader that we have made a
crucial mistake and hardly master the machinery of the expansion of
functions in Tailor series.

After such a successful attack on the psyche of the reader Diakonov
and Petrov have written completely the same expressions up to the replacement $\alpha \to \gamma$ and
$\gamma \to \alpha$ that we have got
by expanding the integrand around the saddle point. In addition they
have added a factor $\cos\beta_n$. We are surprised to find such a
factor, since in fact there should be the factor
\begin{eqnarray}\label{label10.7}
\cos\Bigg(\frac{\beta_n + \beta_{n+1}}{2}\Bigg).
\end{eqnarray}
This is clearly seen from Eq.(89) of our paper [23]. The variables
$\beta_{n+1}$ and $\beta_n$ are independent. The reason to set
$\beta_{n+1} = \beta_n$ would be a mystery of the Diakonov--Petrov
approach to the evaluation of integrals within a saddle point
technique. 

Then, in our paper [23] the factor Eq.(\ref{label10.7}) has been
replaced by unity. The reason of this replacement is in the meaning of
the saddle point technique of the evaluation of integrals. Indeed,
the main point of the saddle point evaluation of integrals is in the
reduction of integrands to the Gaussian form. In the case of the
evolution operator $Z_{\rm Reg}(R_2,R_1)$, as has been claimed by
Diakonov and Petrov in Ref.[3], the saddle point of the integrand
should be at the unit element where $\alpha, \gamma, \beta \ll
1$. That is why the expansions should contain the least powers of
variables. This imposes the exponent of the integrand not to contain
the powers of variables higher than 2. That is why the factor
(\ref{label10.7}) multiplied by $(\alpha_n - \alpha_{n+1})$,
$(\alpha_n - \alpha_{n+1})^2$, $(\gamma_n - \gamma_{n+1})$, $(\gamma_n
- \gamma_{n+1})^2$ should be replaced by unity around the saddle point
$\beta_{n+1}, \beta_n \ll 1$.

As we have shown in our paper [23] the evolution operator $Z_{\rm
Reg}(R_2,R_1)$ is equal to zero (see Eq.(113) of Ref.[23])
\begin{eqnarray}\label{label10.8}
Z_{\rm Reg}(R_2,R_1) = 0.
\end{eqnarray}
This agrees completely with our result $W_{1/2}(C) = W_{1/2}(\rho) =
0$ obtained in this paper. Therefore, the claim "The factors $
cos\beta_n$ being reinstalled, the derivation returns to that of our
paper [3]."  is nothing more than a blef assuming to reanimate the
erroneous result.

Now we would cite again Diakonov and Petrov (see p.12 of Ref.[15]):"
The last objection by FITZ is to our alternative (and in fact
equivalent) derivation of the evolution operator, this time through
the standard Feynman representation for the path integral as a sum
over intermediate states. FITZ quote our result for the evolution
operator of an axial top with the "Wess--Zumino" term, evolving from
its orientation given by a unitary matrix $R_1$ at time $t_1$ to
orientation $R_2$ at time $t_2$:
$$
Z_{\rm Reg}(R_2,R_1) =
(2J+1)\,D^J_{JJ}(R_2R^{\dagger}_1)\,\exp\Bigg[-i(t_2-t_1)\,
\frac{J}{2I_{\perp}}\Bigg]\eqno(39)
$$
where $I_{\perp}$ is a regular moment of inertia, $I_{\perp} \to
0$. Apart from a nontrivial dependence on the orientation matrices
$R{1,2}$ coming through the Wigner $D$--function, this expansion
contains a phase factor $\exp(-i(t_2-t_1)\ldots)$. It is an overall
factor independent of the external field: it can and should be absorbed
into the integration measure to make the evolution operator unity for
the trivial case $R_2 = R_1 = 1$. Indeed, dividing the time interval
into $N$ pieces of small length $\delta$, $N\delta = t_2 - t_1$, one
can write this factor as product,
$$
\exp\Bigg[-i(t_2-t_1)\,\frac{J}{2I_{\perp}}\Bigg] =
\prod^{N}_{k=1}\exp\Bigg[-i\,\delta\,\frac{J}{2I_{\perp}}\Bigg],
\eqno(40)
$$
where, according to the regularization prescription of ref.[1],
$\delta/I_{\perp} \ll 1$ so that each factor is close to unity. Each
factor can be now absorbed into the integration measure $dR(t_k)$ in
the functional--integral representation for the evolution operator
(39). The fact that the factor is complex is irrelevant; moreover, it
is typical for the path--integral representation of the evolution
operators to have a complex measure, see the classical
Feynman$^{\prime}$s book [19]. However, FITZ write:"$\ldots$ a removal
of the fluctuating factor is prohibited since this leads to the change
of the starting symmetry of the system from $SU(2)$ to $U(2)$". It may
seem that FITZ believe that an absorption of a constant factor into
the integration measure changes the number of variables which one
integrates."

It seems that the decomposition of the finite interval of time into
$N$ pieces $t_2 - t_1 = N\,\delta$ is irrelevant to the problem of the
removal of the oscillating factor.  Indeed, the expression given by
the formula (39) does not contain any integration and, therefore,
there is no measure that can absorb the factor
\begin{eqnarray}\label{label10.9}
\exp\Bigg[-i(t_2 - t_1)\,\frac{J}{2I_{\perp}}\Bigg]
\end{eqnarray}
strongly oscillating at $I_{\perp} \to 0$.

In the paper by Faber {\it et al.} the crucial contribution of the
oscillating factor (\ref{label10.9}) has been demonstrated for the example
of the Wilson loop (see Eq.(126) of Ref.[23])
\begin{eqnarray}\label{label10.10}
W^{\rm DP}_J(C) = W_J(C)\,\exp\Bigg[-i(t_2-t_1)\,\frac{J}{2I_{\perp}}\Bigg],
\end{eqnarray}
where $W^{\rm DP}_J(C)$ is the Wilson loop in the Diakonov--Petrov
representation and $W_J(C)$ is the standard Wilson loop defined via
the path--ordering operator. Since the average value of the Wilson
loop is an observable quantity and any averaging over gauge fields
does not affect the oscillating factor Eq.(\ref{label10.9}), one can
write
\begin{eqnarray}\label{label10.11}
\langle W^{\rm DP}_J(C)\rangle = \langle W_J(C)\rangle
\,\exp\Bigg[-i(t_2-t_1)\,\frac{J}{2I_{\perp}}\Bigg].
\end{eqnarray}
According to Wilson's criterion of confinement one should set
\begin{eqnarray}\label{label10.12}
\langle W_J(C)\rangle = e^{\textstyle - \sigma\,{\cal A}},
\end{eqnarray}
where $\sigma$ and ${\cal A}$ are a string tension and a minimal area,
respectively.

Substituting Eq.(\ref{label10.12}) in Eq.(\ref{label10.11}) one arrives
at the expression
\begin{eqnarray}\label{label10.13}
\langle W^{\rm DP}_J(C)\rangle = e^{\textstyle - \sigma\,{\cal A}}
\,\exp\Bigg[-i(t_2-t_1)\,\frac{J}{2I_{\perp}}\Bigg].
\end{eqnarray}
It seems to be rather obvious that the r.h.s. of Eq.(\ref{label10.13})
tends to zero due to the strongly oscillating factor, i.e. 
\begin{eqnarray}\label{label10.14}
\langle W^{\rm DP}_J(C)\rangle_{\rm Reg} &=& \lim_{I_{\perp}\to
0}\langle W^{\rm DP}_J(C)\rangle = \lim_{I_{\perp}\to 0} e^{\textstyle
- \sigma\,{\cal A}}
\,\exp\Bigg[-i(t_2-t_1)\,\frac{J}{2I_{\perp}}\Bigg] =\nonumber\\
&=&e^{\textstyle - \sigma\,{\cal A}} \lim_{I_{\perp}\to 0}
\,\exp\Bigg[-i(t_2-t_1)\,\frac{J}{2I_{\perp}}\Bigg] = 0.
\end{eqnarray}
There is no quantity that can absorb this factor.

Finally, our explanation concerning the crucial influence of the
removal of the oscillating factor (\ref{label10.9}) on the symmetry
of the gauge fields has been clearly given in our paper Ref.[23]. We
would not repeat it here and relegate readers to this publication.

\subsection{Lattice--regularized formula for the Wilson loop 
suggested by Diakonov and Petrov is meaningless}

\hspace{0.2in} In this Section we would to make comments on the
lattice--regularized formula suggested by Diakonov and Petrov in
Ref.[15]. Formula (50) of the manuscript
[15]
$$
W_J ={\cal N}^{-1}\int\prod^N_{k=1}dS_k\,\exp\,\frac{z}{2}\,{\rm
Tr}(S^{\dagger}_kU_{k,k-1}S_{k-1}\tau_3)\eqno(50)
$$
is a revised version of the lattice representation of the Wilson
loop suggested by Diakonov and Petrov in Ref.[7]. However, the
parameter $z$ is now not equal to $z = 2J$, as has been stated in Ref.[7],
since {\it there is no {\bf a priori} reason to expect that in the
lattice regularization $z$ should be the same.}  [15].  Therefore, now
it is suggested to consider this parameter as a free parameter
$z(J) \ne 2J$. We would like to accentuate that expression (50) for
Wilson loops according to Diakonov and Petrov is valid for any
irreducible representation $J$ of gauge group $SU(2)$.

In order to show that the representation Eq.(50) reproduces the
standard Wilson loop
$$
W_J = \frac{1}{2J+1}\,\chi_J(U_{N,N-1}U_{N-1,N-2}\ldots U_{1,N})\eqno(49)
$$
Diakonov and Petrov suggested to expand {\it the exponent} of (50)
according to (34)\footnote{``We make use of the fact that $t_3 =
(-i/2)\tau_3 = (-i/2)\exp(i\pi t_3)$ where the last factor is
definitely an element of $SU(2)$''[15].  It is obvious that the
relation $t_3 = (-i/2)\exp(i\pi t_3)$ is valid only for the matrix
$t_3$ in the fundamental representation $J =1/2$.}:
$$
\exp z{\rm Tr}[t_3U] = \exp\Big\{(-iz/2)\,{\rm Tr}\Big[e^{\textstyle
i\pi t_3}U\Big]\Big\} = \sum_j\tilde{a}_j(z)\,\chi_j\Big[e^{\textstyle
i\pi t_3}U\Big]
$$
$$
=\sum_j\tilde{a}_j(z)\,\sum^{j}_{m = - j}e^{\textstyle i\pi
m}\,D^j_{mm}(U), \quad\tilde{a}_j(z)= e^{\textstyle -i\pi
j}(2j+1)\,\frac{2J_{2j+1}(z)}{z},\eqno(34)
$$
where ``the coefficients $\tilde{a}_j(z)$ being well--known from the
lattice strong--coupling expansion [20]'' [15],
$$
W_J = {\cal N}^{-1}\int
\prod^{N}_{k=1}dS_k\sum_{j_k}\tilde{a}_{j_k}(z)\sum^{j_k}_{m_k = -
j_k}e^{\textstyle i\pi m_k}D^{j_k}_{m_k
m_k}(S^{\dagger}_kU_{k,k-1}S_{k-1}),\eqno(51)
$$
$$
\tilde{a}_{j}(z) = a^{\textstyle -i\pi
j}(2j+1)\,\frac{2}{z}\,J_{2j+1}(z).\eqno(52)
$$
Integrating over the matrices $S_k$ Diakonov and Petrov have arrived at
the expression [15]
$$
W_J = {\cal N}^{-1}\sum_{j}[b_j(z)]^N\chi_j(U_{N,N-1}U_{N-1,N-2}\ldots
U_{1,N}),\eqno(56)
$$
$$
{\cal N} = \sum_{j}(2j + 1)\,[b_j(z)]^N,\eqno(57)
$$
where $b_j(z) = (2/z)\,J_{2j+1}(z)$ [15]. Then, Diakonov and Petrov
claim that at $N \gg 1$ the r.h.s. of (56) reduces to {\it the Wilson
loop in representation $J$} [15].

Now let us adduce some numerical samples demonstrating the erroneous
properties of the representation of the Wilson loop given by Eq.(56).

\noindent{\bf Wilson loop for the fundamental representation $J=1/2$,
$W_{1/2}(z)$}. For the Wilson loop in the fundamental representation
$J=1/2$ the expansion of the exponential in (50) leading to (56)
is valid. The result of the numerical calculation of the coefficients
in (56) and the normalization factor (57) reads
\begin{eqnarray}\label{label10.15}
W_{1/2}(z)|_{z=3.25103} &=& 3.027\times 10^{-8} +
\frac{1}{2.002}\,\chi_{1/2}(U_{24,23}\ldots U_{1,24})\nonumber\\ 
&+& 2.264\times 10^{-4}\,\chi_{1}(U_{24,23}\ldots U_{1,24})+ 4.751\times
10^{-12}\, \chi_{3/2}(U_{24,23}\ldots U_{1,24})\nonumber\\ &+& \ldots
\approx \frac{1}{2}\,\chi_{1/2}(U_{24,23}\ldots U_{1,24}).
\end{eqnarray}
The parameter $z=3.25103$ is taken from Table 1 of Ref.[15]. Thus, for
the fundamental representation the expression (56) fits well the
standard Wilson loop (49).

\noindent{\bf Wilson loop for the adjoint representaion $J=1$,
$W_1(z)$}. First, let us take the Diakonov and Petrov point of view
and believe that the expression (56) is valid for the Wilson loop
defined for the adjoint representation. On this way we have to set $z
= 4.36765$ from Table 1 of Ref.[15] and get the dominant contribution
of the character $\chi_1(U_{24,23}\ldots U_{1,24})$ with the prefactor
$1/3$ according to the normalization of the standard Wilson loop
(49). The numerical analysis gives
\begin{eqnarray}\label{label10.16}
\hspace{-0.5in}W_1(z)|_{z=4.36765} &=& 1.384\times 10^{-9} +
1.795\times 10^{-6}\,\chi_{1/2}(U_{24,23}\ldots U_{1,24})\nonumber\\
\hspace{-0.5in}&+& \frac{1}{3.008}\,\chi_{1}(U_{24,23}\ldots U_{1,24})
+ 6.366\times
10^{-4}\, \chi_{3/2}(U_{24,23}\ldots U_{1,24})\nonumber\\
\hspace{-0.5in} &+& 1.818\times 10^{-10}\,\chi_2(U_{24,23}
\ldots U_{1,24})\ldots
\approx \frac{1}{3}\,\chi_1(U_{24,23}\ldots U_{1,24}).
\end{eqnarray}
This means that really the Diakonov--Petrov representation (56) fits
well the standard Wilson loop (49) defined for the irreducible
representation $J=1$.

\noindent{\bf Wilson loop for the representaion $J=3/2$,
$W_{3/2}(z)$}. Setting $z = 5.46564$ (see Table 1 of Ref.[15]) we
should get in the r.h.s. of (56) the dominant contribution of the
character $\chi_{3/2}(U_{24,23}\ldots U_{1,24})$:
\begin{eqnarray}\label{label10.17}
\hspace{-0.5in}W_{3/2}(z)|_{z=5.46564} &=& 7.145\times 10^{-3} +
5.054\times 10^{-15}\,\chi_{1/2}(U_{24,23}\ldots U_{1,24})\nonumber\\
\hspace{-0.5in}&+& 1.430\times 10^{-5}\,\chi_{1}(U_{24,23}\ldots
U_{1,24})+ \frac{1}{4.051}\, \chi_{3/2}(U_{24,23}\ldots
U_{1,24})\nonumber\\ \hspace{-0.5in}&+& 1.089\times
10^{-3}\chi_2(U_{24,23}\ldots U_{1,24})\ldots \approx
\frac{1}{4}\,\chi_{3/2}(U_{24,23}\ldots U_{1,24}).
\end{eqnarray}
This fits well the standard Wilson loop (49) defined for the
irreducible representation $J=3/2$.

\noindent{\bf Wilson loop for the representaion $J=2$,
$W_2(z)$}. Setting $z = 6.55104$ (see Table 1 of Ref.[15]) we
should get in the r.h.s. of (56) the dominant contribution of the
character $\chi_2(U_{24,23}\ldots U_{1,24})$:
\begin{eqnarray}\label{label10.18}
\hspace{-0.3in}W_2(z)|_{z=6.55104} &=& 1.073\times 10^{-11} +
2.381\times 10^{-3}\,\chi_{1/2}(U_{24,23}\ldots U_{1,24})\nonumber\\
\hspace{-0.3in}&+& 2.392\times 10^{-22}\,\chi_{1}(U_{24,23}\ldots
U_{1,24})+ 5.096\times 10^{-5} \chi_{3/2}(U_{24,23}\ldots
U_{1,24})\nonumber\\
\hspace{-0.3in}&+&\frac{1}{5.073}\chi_2(U_{24,23} \ldots U_{1,24}) 
+ 1.573\times
10^{-3}\chi_{5/2}(U_{24,23}\ldots U_{1,24})\ldots \nonumber\\
 \hspace{-0.5in}&\approx& \frac{1}{5}\,\chi_2(U_{24,23}\ldots
U_{1,24}).
\end{eqnarray}
This fits well the standard Wilson loop (49) defined for the
irreducible representation $J=2$.

\noindent{\bf Wilson loop for the representaion $J=5/2$,
$W_{5/2}(z)$}. Setting $z = 7.62728$ (see Table 1 of Ref.[15]) we
should get in the r.h.s. of (56) the dominant contribution of the
character $\chi_{5/2}(U_{24,23}\ldots U_{1,24})$:
\begin{eqnarray}\label{label10.19}
\hspace{-0.1in}W_{5/2}(z)|_{z=7.62728} &=& 2.051\times 10^{-9} +
3.058\times 10^{-7}\,\chi_{1/2}(U_{24,23}\ldots U_{1,24})\nonumber\\
\hspace{-0.1in}&+& 3.210\times 10^{-4}\,\chi_1(U_{24,23}\ldots
U_{1,24})+ 3.123\times 10^{-38} \chi_{3/2}(U_{24,23}\ldots
U_{1,24})\nonumber\\
\hspace{-0.1in}&+&1.201\times 10^{-4}\chi_2(U_{24,23} \ldots U_{1,24}) 
+ \frac{1}{6.097}\,\chi_{5/2}(U_{24,23}\ldots U_{1,24})\ldots \nonumber\\
 \hspace{-0.1in}&\approx& \frac{1}{6}\,\chi_{5/2}(U_{24,23}\ldots
U_{1,24}).
\end{eqnarray}
This evidences that for $J=5/2$ the Diakonov--Petrov representation of
the Wilson loop (56) fits well the standard Wilson loop (49).

\noindent{\bf Wilson loop for the representaion $J=3$,
$W_3(z)$}. Setting $z = 8.69644$ (see Table 1 of Ref.[15]) we
should get in the r.h.s. of (56) the dominant contribution of the
character $\chi_3(U_{24,23}\ldots U_{1,24})$:
\begin{eqnarray}\label{label10.20}
\hspace{-0.1in}W_3(z)|_{z=8.69644} &=& 6.698\times 10^{-4} +
1.937\times 10^{-17}\,\chi_{1/2}(U_{24,23}\ldots U_{1,24})\nonumber\\
\hspace{-0.1in}&+& 2.681\times 10^{-5}\,\chi_1(U_{24,23}\ldots
U_{1,24})+ 2.807\times 10^{-5} \chi_{3/2}(U_{24,23}\ldots
U_{1,24})\nonumber\\
\hspace{-0.1in}&+&7.627\times 10^{-32}\chi_2(U_{24,23} \ldots U_{1,24}) 
+ 2.216\times 10^{-4}\chi_{5/2}(U_{24,23}\ldots U_{1,24})\nonumber\\
\hspace{-0.1in}&+&\frac{1}{7.158}\,\chi_3(U_{24,23}\ldots
U_{1,24}) + \ldots \approx \frac{1}{7}\,\chi_3(U_{24,23}\ldots
U_{1,24}).
\end{eqnarray}
Thus, the Diakonov--Petrov representation of the Wilson loop (56) fits
well the standard Wilson loop (49) defined for the irreducible
representation $J=3$.

Now let us explain the real meaning of the Diakonov--Petrov
representation for the Wilson loop given by (56). {\bf This
representation is meaningless.}

It is obvious from the following fact. The representation (56) has
been obtained via an expansion valid only for the fundamental
representation $J=1/2$. Therefore, (56) should be written only as
follows
\begin{eqnarray}\label{label10.21}
W_{1/2}(z) &=& {\cal N}^{-1}\sum_{j}[b_j(z)]^N
\chi_j(U_{N,N-1}U_{N-1,N-2}\ldots
U_{1,N}),\nonumber\\
{\cal  N} &=& \sum_{j}(2j + 1)\,[b_j(z)]^N.
\end{eqnarray}
Without discussing the appearance of the artificial normalization
factor ${\cal N}$ we argue that tuning the parameter $z$ we get the
following set of relations
\begin{eqnarray}\label{label10.22}
W^{\rm DP}_{1/2}(z)|_{z=3.25103} &\approx& \frac{1}{2}\,
\chi_{1/2}(U_{N,N-1}U_{N-1,N-2}\ldots U_{1,N}) = W_{1/2},\nonumber\\
W^{\rm DP}_{1/2}(z)|_{z=4.36765} &\approx&
\frac{1}{3}\,\chi_1(U_{N,N-1}U_{N-1,N-2}\ldots U_{1,N}) =
W_1,\nonumber\\ W^{\rm DP}_{1/2}(z)|_{z=5.46564} &\approx&
\frac{1}{4}\,\chi_{3/2}(U_{N,N-1}U_{N-1,N-2}\ldots U_{1,N}) =
W_{3/2},\nonumber\\ W^{\rm DP}_{1/2}(z)|_{z=6.55104} &\approx&
\frac{1}{5}\,\chi_2(U_{N,N-1}U_{N-1,N-2}\ldots U_{1,N}) =
W_2,\nonumber\\ W^{\rm DP}_{1/2}(z)|_{z=7.62728} &\approx&
\frac{1}{6}\,\chi_{5/2}(U_{N,N-1}U_{N-1,N-2}\ldots U_{1,N}) =
W_{5/2},\nonumber\\ W^{\rm DP}_{1/2}(z)|_{z=8.69644} &\approx&
\frac{1}{7}\,\chi_3(U_{N,N-1}U_{N-1,N-2}\ldots U_{1,N}) = W_3,
\end{eqnarray}
where $W^{\rm DP}_{1/2}(z)$ is the Wilson loop in the Diakonov--Petrov
representation defined for the fundamental representation and $W_J$ is
the standard Wilson loop for the irreducible representation $J=1/2, 1,
\ldots$.

{\bf Since {\it a priori} the value of the parameter $z$ is not well
defined, so tuning $z$ we are able to equate the Wilson loop in the
Diakonov--Petrov representation defined only for the fundamental
representation $J=1/2$ to the standard Wilson loop defined for any
irreducible representation $J$.} This is really {\it a new way to
check confinement on lattice} [7,15].

\section*{Instead of Acknowledgement}

\hspace{0.2in} We conclude that in Ref.[15] Diakonov and Petrov have
sold us a conjecture as a proof. In a tedious calculation this
conjecture turned out to be wrong. A mistake in our calculations,
unessential for the drawn conclusions, and some misprints were used to
hide the necessity to redraw the conjecture. New fit parameters were
introduced by them actually modifying the conjecture. The proof of the
new, modified conjecture is again not conclusive for $J \ne
\frac{1}{2}$ or reduces to a triviality. We apologize if we fought too
hard and promise not to recalculate any of Diakonov and Petrov's
papers in near future.

\end{document}